\documentclass[varenna]{cimento}

\def\d{{\rm d}}

\def\d{{\rm d}}

\newcommand{\be}{\begin{equation}}
\newcommand{\ee}{\end{equation}}
\newcommand{\bea}{\begin{eqnarray}}
\newcommand{\eea}{\end{eqnarray}}

\usepackage{graphicx,amsmath,feynmp,amssymb}  
\title{Cosmic ray interactions with matter and radiation}
\author{Pasquale Dario Serpico}
\institute{Laboratoire d'Annecy-le-Vieux de
Physique Th{\'e}orique, USMB, CNRS, F-74940 Annecy, France}

\shortauthor{P.~D.~Serpico}

\begin{document}

\maketitle

\begin{abstract}
The main goal of the present lectures is to outline the key particle interactions and energy loss mechanisms in the Galactic medium that high-energy particles are subject to. These interactions are an important ingredient entering the cosmic ray propagation equation, contributing to shape cosmic ray spectra. They also source the so-called secondary species, like gamma rays, neutrinos, ``fragile'' nuclei not synthesised in stars, and antiparticles, all routinely used as diagnostic tools in a multi-messenger context. These lectures are complementary to Denise Boncioli's ones, focusing instead on processes happening at ultra-high energies in the extragalactic environment. They include prop{\ae}deutic material to Felix Aharonian's and, to some extent, Stefano Gabici's and Carmelo Evoli's lectures.
\end{abstract}

\section{Introduction}   
As illustrated in other courses of these lecture series, in cosmic ray (CR)~\footnote{Here I use the term CRs in a broad sense to mean non-thermal, high-energy astrophysical particles, not necessarily ``measured at the Earth''. While focusing on charged particles, I will also comment on some aspects involving the secondary neutrals produced in their interactions.}  physics we are in general interested in a {\it statistical}  description of the species, typically encoded in ensemble averages (and often angular-averages) of the single-particle distribution functions, sufficient to describe such rarefied species. The corresponding set of coupled {\it Boltzmann equations} is the master tool to describe such systems. Compared to other settings, however, most of the peculiarities of the CR dynamics arise from the highly non-trivial nature of the {\it transport} or left-hand-side (LHS) part of the Boltzmann-Vlasov equation, where  {\it collisionless} phenomena (involving electromagnetic inhomogeneities in the magnetised environments where CRs evolve) are responsible for effects like spatial or momentum diffusion. This does not mean, however, that {\it collisional} effects, i.e. the right-hand-side (RHS) of the Boltzmann equation, are unimportant:  For instance, the maximum energy at which particles can be accelerated, $E_{\rm max}$, or  the spectra and the composition of CRs are influenced by those terms.  Also, charged CRs give rise {\it collisionally} to $\gamma$-rays and $\nu$'s, which are of great importance for astrophysical diagnostics. The structure of this course is the following: In this section we introduce some basic notions and jargon. Sec.~\ref{IonCoul} deals with ionisation and Coulomb losses, while Sec.~\ref{Brems} deals very briefly with Bremsstrahlung: These two sections include the essentials of the electromagnetic CR interactions with interstellar {\it matter}.  Then, in Sec.~\ref{IntEM}, we cover the topic of interactions of CRs with photon and magnetic {\it fields}, which is of importance for electrons and positrons.  Sec.~\ref{hadlosses} deals with hadronic interactions, discussing their role not only as energy loss mechanisms but above all as source terms for {\it secondary} byproducts. The most important species produced as byproducts of both leptonic and hadronic interactions are photons: Sec.~\ref{PhInter} deals with some aspects of high-energy photon interactions, of relevance to some extent for direct detection of gamma rays, but above all for extragalactic very high-energy photons and links with very high-energy neutrinos. It provides thus a natural link with topics covered in D. Boncioli's lectures. Finally, Sec.~\ref{Applications} makes use of a number of the notions outlined, incorporating them in some (simplified) CR propagation problems to substantiate some statements made in this introductory paragraph.

Fundamental microscopic quantities like cross-sections, hopefully familiar from nuclear and particle physics courses, are the  building blocks needed in the considerations that follow.   I will use {\it natural units} for microscopic scales (where $c=k_B=\hbar=1$), so that e.g. energy, mass, momenta, temperature, inverse length and inverse time have the same unit,  eV and multiples. Astrophysical units (parsecs and multiples) are instead used for distances. Cross sections are in barns (1 b=10$^{-24} $cm$^2$). In agreement with most astrophysical literature, the Gaussian electromagnetic convention is used (with the $4\pi$'s in Maxwell equations, not in Coulomb or Biot-Savart laws). The charge of the positron is $e=\sqrt{\alpha}\simeq \sqrt{1/137}\simeq 0.085$. The magnetic field energy density is for instance $B^2/(8\pi)$.

The environmental density of targets and their energy distribution are crucial in determining the  {\it mean-free-path} (mfp) $\ell$ and the associated {\it collision rate} $\Gamma$ of a particle. If $\sigma$ is the cross-section of the interaction process, $\beta$ is the  particle velocity (associated to the Lorentz factor $\gamma=(1-\beta^2)^{-1/2}$), 
its mean-free-path and interaction rate are
\begin{equation}
\ell =\frac{1}{\sigma\, n }, \:\:\:\Gamma=\sigma\,\beta\,n=\frac{\beta}{\ell}\label{mfpir}\,.
\end{equation}
Over an infinitesimal distance interval  $\d s$, the dimensionless {\it optical depth} associated to the process having mfp $\ell$ is $\d\tau =\d s/\ell$. 

Knowing the differential cross-section with respect to the energy transferred $W$, $\d \sigma /\d W$,  one can
define {\it moments of the energy loss}; for instance, the first moment, known as  {\it stopping power}, is
\begin{equation} 
-\frac{{\rm d} E}{{\rm d}x}=-\frac{1}{\beta}\frac{\d E }{\d t}=-\frac{\d |{\bf p}| }{\d t}\equiv n\,\int \d W\, W\frac{ \d \sigma}{\d W}\,.\label{stpower}
\end{equation}
Similarly, the second moment  can be used to define an {\it energy straggling parameter} in terms of which to describe the variance of $E$ losses, see~\cite{Rossi,Jackson}. 
Loss rates 
are useful to estimate the stopping time or range over which the particle of mass $m$ loses its energy via the process:
\begin{equation}
t_{\rm loss}\equiv \int_m^E \frac{\d E'}{-{\rm d} E'/{\rm d}t} \sim \frac{E}{-{\rm d} E/{\rm d}t}\,,\:\:\: d_{\rm loss}\equiv \int_m^E \frac{\d E'}{-{\rm d} E'/{\rm d}x} \sim \frac{E}{-{\rm d} E/{\rm d}x}\,.\label{tauloss}
\end{equation}
Collision are dubbed {\it catastrophic} when the  particle of interest disappears or the fraction of its initial energy $E$ transferred to other particles, $\Delta E/E$, is sizeable.  In this case, $t_{\rm loss}\sim \Gamma^{-1}$ and $d_{\rm loss}\sim \ell$, and typically Monte Carlo methods must be used for a detailed description. Examples in the recent literature are ref.s~\cite{John:2022asa,Esmaeili:2022cpz}. If the primary particle retains its nature and most of its energy in each elementary process,  losses are dubbed {\it continuous}. In this case, $t_{\rm loss}\gg \Gamma^{-1}$ and $d_{\rm loss}\gg \ell$.

In a 
collision,  the square of the center-of-momentum (CoM) energy
is a relativistic invariant:
\begin{equation}
s=\left(\sum_i p_{i}\right)^{2}=
m_{a}^{2}+m_{b}^{2}+2 E_{a} E_{b}\left(1-\beta_{a} \beta_{b} \cos \vartheta\right)\,,\label{sinvariant}
\end{equation}
where the last equality is specific for a two-body collision $a+b\to X$, with $\vartheta$ the angle between the two momenta.
If equated to the square of the sum of masses of the final state, this allows one to estimate threshold energies. For instance, the minimum energy in the Lab frame to produce one antiproton in a collision of a CR proton with a medium proton at rest  can be inferred applying $s$-conservation to  
$pp\to \bar{p}ppp$ (process of the type  $pp\to \bar{p}\,X$ with the lightest $X$ compatible with charge and baryon number conservation), with the final state particles at rest in the Lab frame; eq.~(\ref{sinvariant}) reduces to $2m_p^2+2E_pm_p=(4m_p)^2$, i.e. $E_p>7\,m_p\simeq 6.6\,$GeV.

Another useful invariant in $a+b\to a'+b'$ is the (square) of the momentum transferred
\begin{equation}
q^2=\left(p_{a}-p_{a}'\right)^{2}=\left(p_{b}-p_{b}'\right)^{2}\,.\label{tasfunctpapb}
\end{equation}
For the case of a particle of mass $m$ and momentum $p$ scattering with another  particle, indicating with a prime the outgoing momentum and $\vartheta$ the angle between incoming and outgoing momentum, one has
\begin{equation}
q^2=\left(p-p^{\prime}\right)^{2}=2m^2 - 2 E E^{\prime}\left(1-\beta\beta^{\prime} \cos \vartheta\right) 
\to -4 |{\bf p}|^2\sin^2 \vartheta/2\,,\label{q2}
\end{equation}
where the last step is valid for $E=E'$ or in any case in the relativistic limit~\footnote{Note: Any real photon verifies $q^2=0$; if the scattering process is mediated by the electromagnetic interaction, since $q^2\neq 0$ it means that a {\it virtual} photon  is being exchanged.}. This variable helps us illustrate a key difference between processes of interest for CR and collider studies. The reader is familiar with  the strong angular dependence of Rutherford scattering (if not, appendix~\ref{Ruthscatt} provides a reminder) describing the elastic deflection of a particle of charge $Z_p e$ and mass $m$ with momentum $|{\bf p}|$ (velocity $\beta$) impinging on a heavy target at rest of charge $Z_te$. Its (relativistic) differential cross section is
\begin{equation}
\left.\frac{\d\sigma}{\d\Omega}\right|_{\rm R} = \frac{Z_p^2Z_t^2\alpha^2}{4 |{\bf p}|^2\,\beta^2\sin^4\vartheta/2}\Longleftrightarrow \left.\frac{\d\sigma}{\d\Omega}\right|_{\rm R}= \frac{4Z_p^2Z_t^2\alpha^2 \gamma^2 m^2}{q^4}\,,\label{RuthOmega}
\end{equation}
where the RHS gives the manifestly Lorentz-invariant form. Note how the process is dominated by small-angle scatterings (low $q^2$)~\footnote{The small-angle divergence of the Rutherford cross-section is unphysical since when passing far away from the target, the projectile sees the whole electrically neutral medium. A more physical setting consists in computing the scattering from a Yukawa potential $V=Z_t Z_p \alpha e^{-\mu r}/r$ where the Yukawa mass constant $\mu$ is the inverse of the screening length. The corresponding result, 
equivalent to eq.~(\ref{RuthOmega}) where $q^2 $ is replaced by  $q^2+\mu^2$, is dubbed {\it screened} Rutherford cross section. }. 
On the other hand, if one wants to produce a particle of mass $M$ in a collision, a rough criterion must be $|q^2|>M^2$. So, the bulk of the cross-section (what is most relevant for CR physics in the atmosphere, for instance) is dominated by the  so-called {\it forward} (i.e. $|\vartheta|\simeq 0$) physics;  machines like LHC largely (but not exclusively! See e.g.~\cite{LHCForwardPhysicsWorkingGroup:2016ote}) focus instead on large-angle scatterings, which are associated to large exchanged momenta. Together with the relative rarity of very energetic CRs, which makes hard to study processes with large $s$, this is  the main reason why CRs and high-energy collider physics have largely complementary targets.

\section{Ionisation and Coulomb losses}\label{IonCoul}

Both CR hadrons and leptons can interact electrostatically with a medium containing electrons with density $n_e$. A typical benchmark value of $n_e$ in the {\it interstellar medium} (ISM) of the Milky Way is 1 cm$^{-3}$, distributed in atomic, molecular and ionised hydrogen and helium gas (with traces of heavier elements) as well as in dust grains. While this is a very rarefied medium for terrestrial standards, the large distance and long timescales involved nonetheless imply significant CR energy losses. 

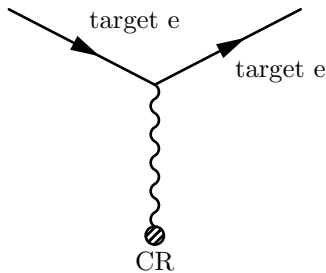
\begin{figure}[!t]
    \begin{center}
               \vspace{0.2cm}      
\begin{fmffile}{fRuth}
    \setlength{\unitlength}{0.6cm}
    \begin{fmfgraph*}(8,5)
       \fmfleft{i1,i2}    
        \fmfright{o1,o2}
        \fmfbottom{b}       
        \fmf{fermion, label=target e, label.side=left}{i2,v1}
        \fmf{fermion,label=target e,label.side=right}{v1,o2}    
        \fmf{photon}{v1,b} \fmfblob{.05w}{b}        
         \fmflabel{CR}{b}
   \end{fmfgraph*}
          \end{fmffile}
    \vspace{0.3cm}      
   \caption{Diagram for ionisation and Coulomb collision losses, whose elementary cross-section can be derived as Rutherford scattering of target electrons in the frame where the CR is at rest.}\label{fig:Rxsec}
  \end{center}
\end{figure}

It is conceptually useful to classify the processes that CR experience depending on the amount of energy $W$ transferred  to their target. If this is below some value $W_{\rm H}$, the motion of the impinging particle of mass $m$ and charge $Ze$ is basically unaffected and the loss can be considered as a continuous process. Technically, this regime is the one where the cross-section is well modelled by {\it Rutherford diffusion} of a fast light projectile in the frame where the CR is at rest (see Fig.~\ref{fig:Rxsec}). From eq.~(\ref{RuthOmega}), changing variable to $q^2$ (note that eq.~(\ref{q2}) implies $\d q^2=-2|{\bf p}|^2 \sin\vartheta \d \vartheta=-|{\bf p}|^2\d\Omega/\pi  $) and accounting for the fact that in the Lab frame, which is the electron rest frame, the energy $W$ transferred to the electron is related to $q$ via $q^2=-2m_e W$, one has 
\begin{equation}
 \left.\frac{\d\sigma}{\d W}\right|_{\rm R} =   \frac{2\pi Z^2\alpha^2}{ m_e \beta^2 W^2}\,,\label{Rutherfordq2}
\end{equation}
 valid between a low energy $W_{\rm L}\gtrsim {\cal O}(B)\,,$ where  $B$  is the binding energy or  related to the  plasma frequency ($B$ is neglected above, since we considered the electrons as free), and  some high value $W_{\rm H}$. Note that in principle the knocked-out electrons also contribute to the low-energy CR population, but this contribution is typically negligible in CR applications~\cite{1966PhRv..150.1088A,1976ApJ...206..312O} . Here we also neglect electron capture and stripping processes, for details see e.g.~\cite{Maurin:2001yxa}.

A posteriori, eq.~(\ref{Rutherfordq2}) justifies neglecting energy losses in interactions with nuclei (so-called {\it nuclear stopping power}), since for a target of mass $M$, $m_e$ is replaced by $ M\gg m_e$ at the denominator of the RHS expression~\footnote{Note that for {\it angular deflection}, instead, nuclei contribute a factor $Z$ more than electrons according to eq.~(\ref{RuthOmega}). This is unimportant for CRs in the ISM,  since spatial transport is collisionless. However, it does matter if one attempts to describe the lateral distribution function of CR showers in the atmosphere, as first successfully tackled by Moli\`ere~\cite{1947ZNatA...2..133M,1948ZNatA...3...78M} and improved upon by Bethe~\cite{Bethe:1953va}.}. 
Using eq.~(\ref{stpower}) one infers
\begin{equation}
-\frac{{\rm d}E}{{\rm d}x}= \frac{2\pi n_e Z^2\alpha^2}{ m_e \beta^2}L_C\,,\:\:\:L_C\simeq\ln \frac{W_{\rm H}}{W_{\rm L}}\label{BB101}\,.
\end{equation}
Eq.~(\ref{BB101}) provides the basic description of the continuous energy losses, with more and more refined considerations replacing  $L_C\simeq 20$, the so-called Coulomb logarithm, with more accurate expressions.  In specific applications to CR physics, relevant formulae---that can be found for instance in~\cite{Mannheim:1994sv,Strong:1998pw,Evoli:2016xgn}---consist in replacing the dimensionless $L_C$  in eq.~(\ref{BB101}) with functions appropriate for the medium conditions and primary particle considered. For conditions of interest to us, note that the $\d E/ \d t $ is almost energy-independent.  

In the following, rather than presenting fitting formulae, let me briefly describe the underlying theoretical approaches through which one completes and improves upon eq.~(\ref{BB101}).
For example, the stopping power integrated at low exchanged energies,  $W< W_{\rm H}$, cannot depend on the arbitrary parameter $W_{\rm L}$. This problem can be solved within a quantum-mechanical treatment,
the first and simplest version of which was provided by Bethe, in terms of a {\it mean excitation potential}, $I$. It yields
\begin{equation}
\left.-\frac{{\rm d}E}{{\rm d}x}\right|_{W<W_H}= \frac{2\pi n_e Z^2\alpha^2}{ m_e \beta^2}\left[\ln\frac{2 m_e \beta^2 W_{\rm H}}{(1-\beta^2)I^2}-\beta^2\right]\,.\label{lowdEdx}
\end{equation}
At the opposite kinematical end when large energies are exchanged, $W>W_{\rm H}$, the Rutherford cross-section becomes in general inadequate, because the description of the projectile (spinless and without recoil in its initial rest-frame) and/or the target electrons (spinless) are insufficiently accurate. For heavy projectiles, often the only correction taken into account is the fact that the electron is a Dirac fermion, not a scalar. That amounts to replace  the Rutherford cross-section by the so-called {\it Mott cross-section}. This  corrects eq.~(\ref{RuthOmega}) by the multiplicative factor $[1-\beta^2\sin^2\vartheta/2]\to \cos^2\vartheta/2$ (the latter in the relativistic limit).   
Kinematics imposes the maximum energy transfer to be  $W_{\rm max}= 2\gamma^2\beta^2 m_e/[1+2 \gamma m_e/m+m_e^2/m^2]\simeq 2\gamma^2\beta^2 m_e$, see Appendix~\ref{WmaxApp} for a derivation. The appropriate modification to eq.~(\ref{Rutherfordq2})  then amounts to the factor $[1-\beta^2W/W_{\rm max}]$. 
Integrating over large energy exchanges, from $W_{\rm H}$ to $W_{\rm max}$  yields
\begin{equation}
\left.-\frac{{\rm d}E}{{\rm d}x}\right|_{W>W_{\rm H}}= \frac{2\pi n_e Z^2\alpha^2}{ m_e \beta^2}\left[\ln\frac{W_{\rm max}}{ W_{\rm H}}-\beta^2\right]\simeq  \frac{2\pi n_e Z^2\alpha^2}{ m_e \beta^2}\left[\ln\frac{2\gamma^2\beta^2 m_e}{ W_{\rm H}}-\beta^2\right]\,.\label{highdEdx}
\end{equation}
Corrections to the Rutherford cross section also arise when accounting for the recoil of the impinging particle and its spin. These lead to  multiplicative factors of unity plus factors suppressed by ratios of the energy to the target mass or powers of them, and are thus negligible for hadronic CRs; further details can be found e.g. in~\cite{Uehling:1954wp}. A brief review on where these ``corrections'' originate from at a fundamental level, as well as the link between the classical Rutherford scattering and the computations of relevant cross-sections in QED is provided in Appendix~\ref{QEDtoRuth}.

Integrating over all transfers $W$, i.e. summing eq.~(\ref{lowdEdx}) to eq.~(\ref{highdEdx}),  finally gives
\begin{equation}
\left.-\frac{{\rm d}E}{{\rm d}x}\right|_{\rm Bethe}\simeq \frac{4\pi n_e Z^2\alpha^2}{ m_e \beta^2}\left [\ln\left( \frac{2m_e\beta^2\gamma^2}{I}\right)-\beta^2\right]\label{BB}\,.
\end{equation}
The microphysics entering the probably familiar {\it Bethe stopping power formula},  eq.~(\ref{BB}), (for a more extensive discussion see e.g. chapter 13 of~\cite{Jackson}) is sufficient to describe losses of hadronic CRs. More accurate descriptions used for laboratory applications often require a number of corrections in the description of the medium and projectile~\cite{Ahlen:1980xr}. For instance, at low-energies the atomic shell structure is important ({\it $C$-correction}), and a projectile with $Z>1$ may not be fully ionised, thus motivating the notion of {\it effective charge}, $Z_{\rm eff}$. At high-energy, medium-polarisation effects are important ({\it density correction}, or $\delta$-term). $Z$-dependent higher-order corrections are also sometimes introduced, e.g. to account for Coulomb distortion of wavefunctions (eq.~(\ref{BB}) assumes Born approximation).  
The free parameters entering these improved descriptions are often empirically-deduced; further details can be found e.g. in~\cite{Ahlen:1980xr}
 or in the ``Passage of particles through matter'' mini-review in the PDG~\cite{ParticleDataGroup:2020ssz}.

\begin{figure}[!t]
\begin{center}
               \vspace{0.2cm}      
\begin{fmffile}{fel}
    \setlength{\unitlength}{0.6cm}
    \begin{fmfgraph*}(8,5)
        \fmfleft{i1,i2}
        \fmfright{o1,o2}
        \fmf{fermion}{i1,v1,o1}
        \fmf{fermion}{i2,v2,o2}
        \fmf{photon}{v1,v2}
          \fmflabel{$e^-$}{i1}
         \fmflabel{$e^-$}{i2}
          \fmflabel{$e^-$}{o1}
         \fmflabel{$e^-$}{o2}
   \end{fmfgraph*}
   \hspace{0.2cm}
      \begin{fmfgraph*}(8,5)
\fmfleft{i1,i2}
        \fmfright{o1,o2}
        \fmf{fermion}{i1,v1}
        \fmf{phantom}{v1,o1} 
        \fmf{fermion}{i2,v2}
        \fmf{phantom}{v2,o2} 
        \fmf{photon}{v1,v2} 
\fmf{fermion,tension=0}{v1,o2} 
\fmf{fermion,tension=0}{v2,o1}
          \fmflabel{$e^-$}{i1}
         \fmflabel{$e^-$}{i2}
          \fmflabel{$e^-$}{o1}
         \fmflabel{$e^-$}{o2}
    \end{fmfgraph*} 
          \end{fmffile}
              \vspace{0.3cm}   
   \caption{Diagrams for M{\o}ller (or electron-electron) scattering: $t$-channel to the left, $u$-channel to the right. There is a minus sign between the two amplitudes, due to the fermionic nature.}\label{fig:Moexsec}
  \end{center}
\end{figure}
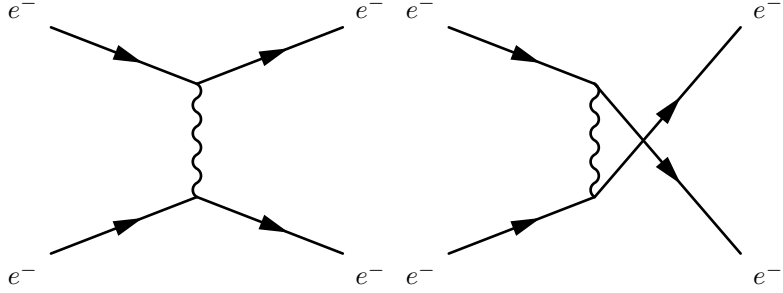
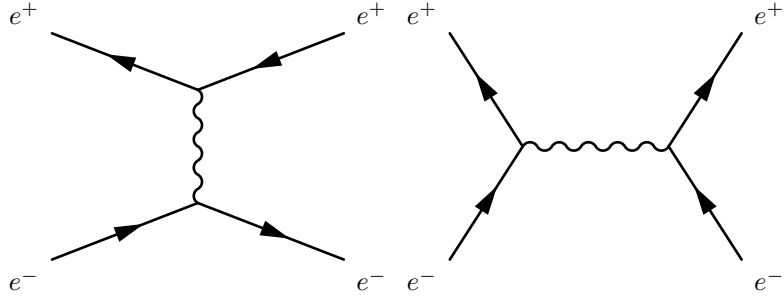
\begin{figure}[!t]
\begin{center}
              \vspace{0.2cm}   
\begin{fmffile}{fpos}
    \setlength{\unitlength}{0.6cm}
    \begin{fmfgraph*}(8,5)
        \fmfleft{i1,i2}
        \fmfright{o1,o2}
        \fmf{fermion}{i1,v1,o1}
        \fmf{fermion}{o2,v2,i2}
        \fmf{photon}{v1,v2}
          \fmflabel{$e^-$}{i1}
         \fmflabel{$e^+$}{i2}
          \fmflabel{$e^-$}{o1}
         \fmflabel{$e^+$}{o2}
   \end{fmfgraph*}
   \hspace{0.2cm}
    \begin{fmfgraph*}(8,5)
        \fmfleft{i1,i2}
        \fmfright{o1,o2}
        \fmf{fermion}{i1,v1,i2}
        \fmf{fermion}{o1,v2,o2}
\fmf{photon}{v1,v2}
          \fmflabel{$e^-$}{i1}
         \fmflabel{$e^+$}{i2}
          \fmflabel{$e^-$}{o1}
         \fmflabel{$e^+$}{o2}
    \end{fmfgraph*}    
          \end{fmffile}
              \vspace{0.3cm}   
   \caption{Diagrams for Bhabha (or electron-positron) scattering: $t$-channel to the left, $s$-channel to the right. There is a minus sign between the two amplitudes, due to the fermionic nature.}\label{fig:Bhxsec}
  \end{center}
\end{figure}

For electrons and positrons, the cross-sections leading to eq.~(\ref{highdEdx}) are inadequate, at least if one wants to describe processes at better than the 10\% level. The recoil of the particles is relevant, the identical nature of particles enters for $e^-$'s, spin does matter and in general more than one Feynman diagram contribute even at tree level.
The relevant $e^-e^-$ scattering   in QED is described by the {\it M{\o}ller} cross section~\cite{Moeller31}\footnote{The Danish physicist Christian M{\o}ller actually inferred this formula heuristically. Its derivation within QED is due to Bethe and Fermi~\cite{1932ZPhy...77..296B}. This may be the appropriate occasion to remind the readers about the often-forgotten role that Fermi played in the early days of QED. Albeit less well-known than many other of his theoretical contributions (quantum statistics, weak theory\ldots), it was crucial especially since he provided an appealing, accessible and coherent synthesis of the advances that colleagues like Dirac, Heisenberg, Pauli and others had obtained. His 1932 pedagogically remarkable review~\cite{1932RvMP....4...87F} was eventually at the basis of the reference textbook by Heitler~\cite{Heitler:1936jqw}. Fermi's approach trained in QED generations of physicists in the 30's and 40's,  influencing people like Feynman and Weisskopf. For a review, see~\cite{Schweber02}.} , see Fig.~\ref{fig:Moexsec}
\begin{equation}
\left.\frac{\d \sigma}{\d W}\right|_{\rm M\o}=\frac{2 \pi \alpha^{2}}{m_e \beta^{2}W^2} \left[1+\frac{W^2}{(E-W)^2}+\frac{(\gamma-1)^2W^2}{\gamma^2E^2}-\frac{2 \gamma-1}{\gamma^{2}}\frac{W}{E-W}\right]\,,
\end{equation}
with the further condition that $W_{\rm max}=E/2$ due to the identical nature of particles.
For $e^+e^-$ scattering,  {\it Bhabha} cross section~\cite{Bhabha:1936zz} is the relevant one, see Fig.~\ref{fig:Bhxsec} for the governing diagrams at tree level:
\begin{eqnarray}
&&\left.\frac{\d \sigma}{\d W}\right|_{\rm Bh}=\frac{2 \pi \alpha^{2}}{m_e \beta^{2}W^2} \left\{1-\frac{\gamma^{2}-1}{\gamma^{2}} \frac{W}{E}+\frac{1}{2}\left(\frac{\gamma-1}{\gamma}\right)^{2}\left(\frac{W}{E}\right)^{2}\right.\\
&&\nonumber-\frac{\gamma-1}{\gamma+1} \frac{W}{E}\left[\frac{\gamma+2}{\gamma}-2 \frac{\gamma^{2}-1}{\gamma^{2}} \frac{W}{E}+\left(\frac{\gamma-1}{\gamma}\right)^{2}\left(\frac{E}{W}\right)^{2}\right] \\
&&\nonumber\left.+\left(\frac{\gamma-1}{\gamma+1}\right)^{2}\left(\frac{W}{E}\right)^{2}\left[\frac{1}{2}+\frac{1}{\gamma}+\frac{3}{2 \gamma^{2}}-\left(\frac{\gamma-1}{\gamma}\right)^{2} \frac{W}{E}\left(1-\frac{W}{E}\right)\right]\right\}\,.
\end{eqnarray}

Relevant stopping power formulae can be found e.g. in~\cite{Uehling:1954wp,ParticleDataGroup:2020ssz}, and are reported here:
 
\begin{equation}
\left.-\frac{{\rm d}E}{{\rm d}x}\right|_{W>W_{\rm H}}^{e^-}= \frac{2\pi n_e\alpha^2}{ m_e \beta^2}\left[\ln\frac{m_e(\gamma-1)}{4 W_{\rm H}}+1-\frac{2\gamma -1}{\gamma^2}\log 2+\frac{1}{8}\left(\frac{\gamma-1}{\gamma}\right)^2\right]\,,\label{highdEdxEl}
\end{equation} 
and
\begin{equation}
\left.-\frac{{\rm d}E}{{\rm d}x}\right|_{W>W_{\rm H}}^{e^+}= \frac{2\pi n_e\alpha^2}{ m_e \beta^2}\left[\ln\frac{m_e(\gamma-1)}{W_{\rm H}}-\frac{\beta^2}{12}\left\{11+\frac{14}{\gamma+1}+\frac{10}{(\gamma+1)^2}+\frac{4}{(\gamma+1)^3}\right\}\right]\,.\label{highdEdxPos}
\end{equation} 
Adding the low energy transfer contribution, one gets:
\begin{equation}
\left.-\frac{{\rm d}E}{{\rm d}x}\right|^{e^-}= \frac{2\pi n_e\alpha^2}{ m_e \beta^2}\left[\ln\frac{m_e^2\beta^{2}\gamma^2(\gamma-1)}{2 I^2}+(1-\beta^2)-\frac{2\gamma -1}{\gamma^2}\log 2+\frac{1}{8}\left(\frac{\gamma-1}{\gamma}\right)^2\right]\,,\label{dEdxEl}
\end{equation} 
and
\begin{equation}
\left.-\frac{{\rm d}E}{{\rm d}x}\right|^{e^+}= \frac{2\pi n_e\alpha^2}{ m_e \beta^2}\left[\ln\frac{2m_e^2\beta^{2}\gamma^2(\gamma-1)}{I^2}-\frac{\beta^2}{12}\left\{23+\frac{14}{\gamma+1}+\frac{10}{(\gamma+1)^2}+\frac{4}{(\gamma+1)^3}\right\}\right]\,.\label{dEdxPos}
\end{equation} 

In addition, positron annihilation should be taken into account at sufficiently low energies; it is obviously responsible for a catastrophic loss. Its cross section is:
\begin{equation}
\sigma_{\rm ann}^{e^+}= \frac{\pi \alpha^2 \left\{\left(\gamma^{2}+4 \gamma+1\right) \ln \left[\gamma+\left(\gamma^{2}-1\right)^{1 / 2}\right]-(3+\gamma)\left(\gamma^{2}-1\right)^{1 / 2}\right\}}{m_e^2(\gamma+1)\left(\gamma^{2}-1\right)} \,.
\end{equation}

Finally, note that the processes described here from the CR perspective are important also for the {\it interstellar medium} (ISM), since they ionise and heat it. Notably, CRs are thought to be the only ionising medium effective in penetrating dense molecular clouds, which are stellar formation nurseries. This is another role played in astrophysics by CR collisional effects, albeit its detailed treatment also requires modeling of atomic and molecular physics not discussed here. For the interested readers, a seminal paper is~\cite{Padovani:2009bj}, while a recent review of the subject can be found in~\cite{Gabici:2022nac}.

\section{Bremsstrahlung}\label{Brems}
\begin{figure}[!t]
\begin{center}
\begin{fmffile}{fbrems}
    \setlength{\unitlength}{0.6cm}
    \begin{fmfgraph*}(8,5)
        \fmfleft{i1,i2}
        \fmfright{o1,o2,o3}
        \fmf{fermion, label=target}{i1,v1,o1}
        \fmf{fermion, label=$e$}{i2,v2,v3,o2}
        \fmf{photon}{v1,v2}
         \fmf{photon, label=$\gamma$}{v3,o3}
   \end{fmfgraph*}
          \end{fmffile}
   \caption{One of the  QED tree-level diagrams for bremsstrahlung.}\label{fig:brems}
  \end{center}
\end{figure}
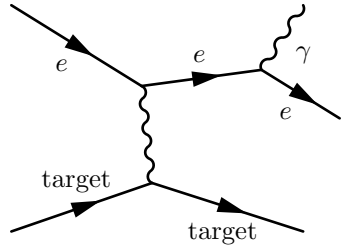

Sufficiently energetic particles tend to {\it radiate} a photon under the braking action of the nuclear and electronic electric fields, a process associated to significant losses for CR leptons, see Fig.~\ref{fig:brems}. 
The relevant differential {\it bremsstrahlung cross section} for an electron of initial energy $E$ to produce a photon of wavenumber $k$ (ending up with energy  $E_f=E-k$) impinging onto the target $i$ can be written as
\begin{equation}
\frac{{\rm d}\sigma_i}{{\rm d}k}=\frac{\alpha^3}{m_e^2k}\left[\left(1+\frac{E_f^2}{E^2}\right)\phi_1^{(i)}-\frac{2}{3}\frac{E_f}{E}\phi_2^{(i)}\right]\,,\label{sigbrems}
\end{equation}
a result due to Bethe and Heitler~\cite{BH34} which relies on Born perturbation theory (plane waves for the incoming/outgoing electron), on whose general form we will comment below.
If the target is an unshielded charge $Ze$,  one has
\begin{equation}
\phi_1=\phi_2=Z^2\phi_u\,,\:\:\:\phi_u=4 \left(\ln \frac{2EE_f}{m_e\,k}-\frac{1}{2}\right)\,,
\end{equation} 
while more complicated expressions hold e.g. for partially shielded nuclei. For a neutral plasma of electrons and a single nucleus, summing eq.~(\ref{sigbrems}) over the two species yields 
\begin{equation}
\frac{{\rm d}\sigma_{\rm tot}}{{\rm d}k}=\frac{4(Z^2+Z)\alpha^3}{m_e^2k}\left[1+\frac{E_f^2}{E^2}-\frac{2}{3}\frac{E_f}{E}\right]\left(\ln \frac{2EE_f}{m_e\,k}-\frac{1}{2}\right)\,.\label{unscreened}
\end{equation}
For scattering on a target of density $n_t$, we thus derive
\begin{equation}
-\left.\frac{{\rm d}E}{{\rm d}x}\right|_{\rm Brems}= n_{\rm t}E\Phi^{\rm rad}_t\,,\:\:\:\Phi^{\rm rad }_t\equiv \frac{1}{E}\int_0^{E} \d k\, k\frac{ \d \sigma}{\d k}\,,\label{dEdxbrems}
\end{equation}
with $\Phi_t^{\rm rad }\sim{\cal O}(\alpha^3)$ and almost (exactly, according to eq.~(\ref{unscreened})) independent on $E$. A few comments worth making:
\begin{itemize} 
\item Compared to eq.~(\ref{Rutherfordq2}), eq. (\ref{sigbrems})  is suppressed by a factor $\alpha$, which finds an obvious interpretation in QED since the elementary diagram for bremsstrahlung contains an extra vertex associated to the radiated photon, compare Fig.~\ref{fig:brems} with Fig.~\ref{fig:Moexsec}.
\item The  bremsstrahlung stopping power, eq.~(\ref{dEdxbrems}), grows in importance with respect to eq.~(\ref{BB}) when $E$ grows, with the stopping timescale for bremsstrahlung defined in eq.~(\ref{tauloss}) nearly independent of $E$. 
\item The analogous of eq.~(\ref{sigbrems}) for CR nuclei  of mass $m=A\,m_p$ and charge $Z\,e$ would lead to a cross section suppressed by $Z^4A^{-2}(m_e/m_p)^2\simeq 10^{-7}(Z^4/A^{2})$, which immediately shows how much more inefficient electromagnetic radiative processes are expected to be for nuclei;  energy losses of CR nuclei at intermediate and high-$E$ are in fact dominated by hadronic interactions, dealt with in Sec.~\ref{hadlosses}.
\end{itemize}

This is but a very short introduction to the topic, which in fact has vast applications in astrophysics and where it is sometimes known also under the name {\it free-free} emission.  Although a quantum treatment is needed for a throughout analysis, several regimes are also amenable to classical approximations. Besides the textbook~\cite{RLrpap91}, another useful and  modern reference adopting the classical electrodynamics approach, with several astrophysical applications, is~\cite{Ghisellini:2012hs}.  A classical reference with a more particle physics approach for the material covered here is~\cite{Blumenthal:1970gc}; a complete reference textbook covering clearly classical and quantum aspects, relativistic and non-relativistic limits is~\cite{Gouldbook}. These are useful references that cover not only the material introduced in this section, but also the one outlined in the following one.

\section{Interactions with electromagnetic fields}\label{IntEM}
Besides rarefied gas (and dust), the ISM of our Galaxy, as well as external ones, also contains photons and is threaded with magnetic fields. These are further targets for energy-loss mechanisms of CRs, described in the following.  

That the ISM is magnetised is most clearly revealed by radio observations (Faraday rotation, synchrotron radiation) but also via polarized light emission (due to dust grains) or, in some cases, via Zeeman splitting.  A magnetic field coherent (at least) up to scales comparable to several kpc-long Galactic structures, such as spiral arms, has been detected. Radio observations of external Galaxies as well as our own, notably via synchrotron emission (see below), indicate several kpc thick magnetized halos embedding the stellar disks. The ISM magnetic field extends to small scales down to at least the typical pc-scale distance between neighbouring stars, where the field orientation is believed to fluctuate following the turbulence of the ISM.  Inferred intensities of Galactic fields (typically via radio synchrotron emission, its polarization and its Faraday rotation) are in the 1-10$\mu$G range. For a relatively recent review and ample references on Galactic and cosmic magnetism, including microphysics and simulations, see for instance~\cite{Akahori:2017lhe}.

The most important photon backgrounds for CR studies are: 
\begin{itemize} 
\item 
The cosmic microwave background or CMB (blackbody of cosmological origin and temperature of about 2.7 K), pervading the whole universe, Galaxy and extragalactic sky alike. While today its energy density is $\sim 0.3\,$eV/cm$^3$, it scales with redshift as $(1+z)^4$. 
\item The extragalactic background light (EBL), pervading the whole universe in the infrared (IR) to ultraviolet (UV) range, due to primary stellar emission and secondary radiation from reprocessing, with a benchmark number density of roughly a photon /cm$^3$; a recent determination of its spectral energy density can be found in~\cite{Ajello:2018sxm}. Backgrounds at different wavelengths also exist (e.g. radio) but are less important for what follows. 
\item In the Galactic environment, in addition there are important UV, optical, and IR backgrounds associated to stellar light emission and its reprocessing, with typically larger energy density than the CMB and  non-homogeneously distributed, notably peaking towards the inner Galaxy. The energy densities of magnetic fields, radiation and cosmic rays are equal within less than one order of magnitude in our Galaxy, amounting to $\lesssim 1$eV/cm$^3$ each.
\end{itemize}

In the ISM of our Galaxy, the benchmark density of  matter $\sim$1 cm$^{-3}$ is to be compared with $\sim $ 410 cm$^{-3}$ CMB photons, and two orders of magnitude lower densities for starlight.  Combined with typical Myr propagation timescales and the much larger hadronic cross sections than electromagnetic ones, this makes obvious that interactions with radiation are irrelevant for CR nuclei propagating in the Galaxy. By the way, 
a CR proton with energy $E=\gamma m_p$  sees a background photon of energy $\epsilon$ as having an energy $E_\gamma\sim \gamma \epsilon$ in its rest-frame: At the typical GeV-TeV energies of interest, and up to the PeV scale, the CR sees  sub-MeV photons, a kinematical factor  preventing potentially interesting inelastic channels to be open. In what follows, we will thus concentrate on interactions of CR electrons with radiation, which are instead relevant also at GeV-TeV energies.

Note that the situation is different for ultra-high energy cosmic rays, whose extragalactic origin implies:
\begin{itemize} 
\item  Propagation times $\gtrsim$ Gyrs.
\item A medium with $\sim 10^7$ times lower densities of matter, making hadronic interactions irrelevant.
\item Larger Lorentz factor ($\gtrsim10^{9}$), thus allowing interesting kinematical threshold for interactions with radiation to be open. 
\end{itemize}
For a treatment of these processes, we refer to D. Boncioli's lectures.

\subsection{Inverse Compton}

\begin{figure}
\begin{center}
    \begin{fmffile}{fcompt}
      \setlength{\unitlength}{0.6cm}
      \begin{fmfgraph*}(8,5)
        \fmfleft{i1,i0}
        \fmfright{o1,o0}
      \fmf{fermion}{w1,w2}
     \fmf{boson,label=$k$}{i0,w1}
      \fmf{boson,label=$k'$}{w2,o0}
      \fmf{fermion,label=$p$}{i1,w1}
      \fmf{fermion,label=$p'$}{w2,o1}
          \end{fmfgraph*}    
          \end{fmffile}
     \begin{fmffile}{fsynch}
      \setlength{\unitlength}{0.6cm}
      \begin{fmfgraph*}(8,5)
        \fmfleft{i1,i0}
        \fmfright{o1,o0}
      \fmf{fermion}{w1,w2}
     \fmf{boson,label=$B$}{i0,w1}
      \fmf{boson,label=$\gamma$}{w2,o0}
      \fmf{fermion,label=$e$}{i1,w1}
      \fmf{fermion,label=$e$}{w2,o1}
       \fmfv{decor.shape=cross}{i0}     
          \end{fmfgraph*}    
          \end{fmffile}         
  \caption{Left: one of the two QED tree-level diagrams for (Inverse) Compton scattering, with labelled momenta. Right: The analogous diagram for synchrotron radiation, with an interaction with a virtual photon associated to an external $B$-field.}\label{fig:BasicQED}
  \end{center}
\end{figure}
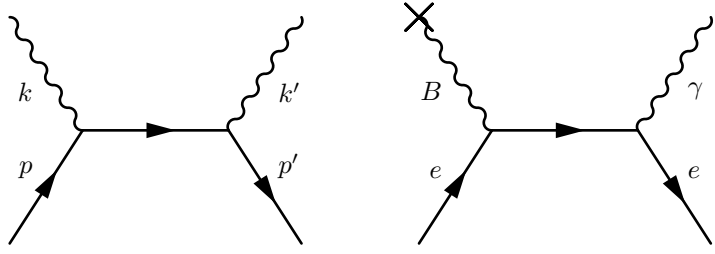

The elementary QED process at the hearth of the interactions of leptons with radiation is the two body scattering $e\gamma\to e'\gamma'$, dubbed {\it Compton scattering} if considering the photon as projectile and the electron as target (at rest), or {\it Inverse Compton} if the electron is the energetic particle transferring part of its energy to a low energy photon (see Fig.~\ref{fig:BasicQED}, left).

First,  let us focus on its kinematics. Imposing energy momentum balance in the photon (whose momenta are labelled via $k,\:k'$)-electron (momentum $p,\:p'$) scattering:
\begin{equation}
k_\mu+p_\mu=k'_\mu+p'_\mu\,,
\end{equation}
 the on-shell condition for the electron mass, $m_e^2=p^2=p'^2$, and that $k^2=k'^2=0$ for real photons, one has:
\begin{equation}
m_e^2=(p+k-k')_\mu(p+k-k')^\mu \Longrightarrow p\cdot (k- k')=k\cdot k'\,.\label{Ckin}
\end{equation}
Let us evaluate eq.~(\ref{Ckin}) in the electron rest-frame, $p=(m_e,0,0,0)$, frame variables that we will denote with tilde; let the $x$-axis be the incoming direction of the photon, and the $x-y$ plane the one spanned by the incoming-outgoing direction. Then
 $k=\tilde\epsilon(1,1,0,0)$, while  $k'=\tilde\epsilon'(1,\cos \tilde\theta,\sin \tilde\theta,0)$, so that
 \begin{equation}
m_e (\tilde\epsilon-\tilde\epsilon')=\tilde\epsilon\tilde\epsilon'(1-\cos\tilde\theta) \Longrightarrow \tilde\epsilon'=\frac{\tilde\epsilon}{1+\frac{\tilde\epsilon}{m_e}(1-\cos\tilde\theta)}\,.\label{relCom}
\end{equation}
Eq.~(\ref{relCom}) illustrates that the ratio $x\equiv \tilde\epsilon/m_e$ controls the photon energy change, so that  unless the energy of the photon  in the electron rest-frame is comparable to or larger than  the mass of the electron $m_e$, the photon energy is only slightly altered. 
If the electron has a velocity $\beta$ (Lorentz factor $\gamma$)  in the Lab frame,  eq.~(\ref{relCom}) is valid in the electron rest-frame, denoted  via a tilde, which is boosted by $\gamma$ with respect to the Lab, so that 
with $\tilde\epsilon=\epsilon\gamma(1-\beta\cos\theta)$. If we want to express $\tilde\epsilon'$ in the Lab frame, we need to perform the ``reverse boost'', such that $\epsilon'=\gamma(1+\beta\cos(\pi-\tilde \theta))\tilde\epsilon'$; hence
\begin{equation}
\epsilon'=\gamma(1-\beta\cos\tilde\theta)\frac{\tilde\epsilon}{1+\frac{\tilde\epsilon}{m_e}(1-\cos\tilde\theta)}=\gamma^2\epsilon\frac{(1-\beta\cos\tilde\theta)(1-\beta\cos\theta)}{1+\frac{\gamma\epsilon}{m_e}(1-\beta\cos\theta)(1-\cos\tilde\theta)}\,.\label{relatCS}
\end{equation}
In the limit where  $\epsilon\,\gamma \ll m_e$, i.e. $\epsilon\,E_e\ll m_e^2$, known as {\it Thomson regime}, with the appropriate aberration relation, $\cos \theta=(\cos\tilde\theta-\beta)/(1-\beta\cos\tilde\theta)$,  one has that 
\begin{equation}
\epsilon'= \gamma^2\epsilon(1-\beta\cos\tilde\theta)^2\Rightarrow \langle \epsilon'\rangle=\frac{4}{3}\gamma^2\epsilon\simeq 5\left(\frac{\epsilon}{\rm eV}\right)\left(\frac{E_e}{\rm GeV}\right)^2\:{\rm MeV}\label{Then}\,,
\end{equation}
where we used the isotropy of the scattered photons in the electron rest-frame to perform the average.
For  $\epsilon\,E_e> m_e^2$, known as {\it Klein-Nishina regime}, one has instead
\begin{equation}
\epsilon'\simeq \gamma m_e\sim E_e\,.\label{KNen}
\end{equation}

These two kinematical regimes also manifest differences in the cross-section of the process.  The general differential cross-section at tree level in QED was derived by Klein and Nishina~\cite{KN29}, constituting one of the first QED predictions (see the diagram in Fig.~\ref{fig:BasicQED}, left); for unpolarised photons in the rest-frame of the electron, it reads
\begin{equation}
\frac{\d \sigma}{\d \Omega}=\frac{\alpha^2}{2m_e^2}\left(\frac{\tilde\epsilon'}{\tilde\epsilon}\right)^{2}\left(\frac{\tilde\epsilon}{\tilde\epsilon'}+\frac{\tilde\epsilon'}{\tilde\epsilon}-\sin ^{2}\tilde \theta\right)\,,
\end{equation}
where $\tilde\epsilon$ ($\tilde\epsilon'$) is the initial (final) energy of the photon, and $\tilde\theta$ its deflection angle. In terms of $\sigma_T\equiv 8\pi \alpha^2/(3m_e^2)$ ({\it Thomson cross-section}) the total cross section writes
\begin{eqnarray}
\sigma &=&2 \pi \int_{0}^{\pi} \frac{\d \sigma}{\d \Omega} \sin \tilde\theta \d \tilde\theta = \nonumber\\
&&\frac{3}{4} \sigma_{T}\left[\frac{1+x}{x^{3}}\left(\frac{2 x(1+x)}{1+2 x}-\ln (1+2 x)\right)+\frac{1}{2 x} \ln (1+2 x)-\frac{1+3 x}{(1+2 x)^{2}}\right] \,,
\end{eqnarray}
so that
\bea
\sigma(x)  &\simeq& \sigma_{T}(1-2 x+\ldots) \:\:\: {\rm for } \:x \ll 1\:  {\rm (Thomson\: limit) } \\ 
\sigma(x) &\simeq& \frac{3}{8} \sigma_{T} \frac{1}{x}\left(\ln 2 x+\frac{1}{2}\right) \:\:\: {\rm for } \: x \gg 1 \:  {\rm (Klein-Nishina\: limit) }\,. 
\eea
Combined with the previously derived kinematical relations eqs.~(\ref{Then},\ref{KNen}), we can thus conclude that in the Thomson limit, collisions are frequent,  the background photon energy gets boosted quadratically in the electron energy, but the electron only loses a small amount of energy, so that the loss can be considered continuous. On the other hand, in the Klein-Nishina limit the cross-section is suppressed, but once interacting the electron transfers a significant fraction of its energy to the background photon. 
Knowing $W=\epsilon'-\epsilon$ and the differential cross section, we can compute the energy loss via eq.~(\ref{stpower}), after changing variable from $\cos\tilde\theta$ to $W$. It is however instructive to deduce the stopping power in the Thomson-limit  with classical considerations.

\subsection{Classical treatment}
In classical electromagnetism, the infinitesimal power ${\rm d}P$ per unit solid angle carried by waves across the infinitesimal surface $\d\vec{A}$ (its direction being the normal $\hat{ {\bf n}} $  to the surface) can be written as
\begin{equation}
{\rm d}P={\bf S}\cdot \d\vec{A}={\bf S}\cdot \hat{ {\bf n}} R^2{\rm d\Omega}\,,
\end{equation}
$R$ being the distance from the radiating source, in terms of the Poynting vector ${\bf S}$ defined as
\begin{equation}
{\bf S}=\frac{{\bf E}\times {\bf B}}{4\pi}=\frac{|{\bf E}|^2}{4\pi}\hat{{\bf S}}\,.\label{Poynting}
\end{equation}
The Poynting vector thus represents the power per unit solid angle carried by waves across the surface normal to their propagation direction.
In the non-relativistic limit, the radiated power  $P$ (which is a $-{\rm d}E/{\rm d}t$) of a charge $q$ accelerating with acceleration $a$ is described by the {\it Larmor formula }
\begin{equation}
\frac{{\rm d}P}{{\rm d}\Omega}=\frac{q^2 a^2\sin^2\phi}{4\pi}\Rightarrow P=\frac{2}{3}q^2 a^2\,,\label{larmor}
\end{equation}
$\phi$ being the angle between the normal to the surface and  the acceleration vector. 
In fact, eq.~(\ref{larmor}) is also valid in the relativistic case, provided that $a^2$ is now replaced by the four-acceleration (or, four-force over mass) squared, $a^2\to a_\mu a^\mu\,, m\,a^\mu=\d p^\mu/\d \tau$.
It is important to remember that the radiated power is a Lorentz-invariant, as one can infer from the fact that $\d E$ and $\d t$ are both the $0$-th components of (parallel) four-vectors.

Consider the interaction of an electromagnetic wave with a charged particle at rest, and let us compute the power that this particle radiates. 
The Lorentz force for $\beta\to 0$  gives an acceleration $q{\bf E}/m$. For a sinusoidal wave $E=E_0\sin(\omega t +\phi)$ along some direction, the Larmor formula gives the (time-averaged) emitted power
\begin{equation}
\langle P \rangle=\frac{2}{3}q^2 \langle a^2\rangle=\frac{2}{3}\frac{q^4}{m^2}\frac{E_0^2}{2}\,.
\end{equation}
In scattering theory, the cross-section is the ratio of the radiated power to incident flux, 
\begin{equation}
\sigma =\frac{\langle P \rangle}{|\langle{\bf S}\rangle|}=\frac{8\pi}{3}\frac{q^4}{m^2}= \sigma_T\:\:\:({\rm if}\:q=\pm e,\:m=m_e )\,,
\end{equation}
where the Poynting vector is $|\langle{\bf S}\rangle|=E_0^2/(8\pi)$ and we used the fact that the electric and magnetic field of the wave have the same amplitude.

We can also write 
\begin{equation}
\langle P \rangle=\sigma_T\,\tilde u\,,\:\:\: \tilde u=\left\langle \frac{|{\bf E}|^2}{8\pi}+\frac{|{\bf B}|^2}{8\pi} \right\rangle=\frac{E_0^2}{8\pi}\,,\label{PowerIC}
\end{equation}
in terms of  $\tilde u$, the energy density in the e.m. field in the electron rest frame.
In a ``particle'' interpretation, the power in eq.~(\ref{PowerIC})  can be interpreted as equal to the scattering rate (itself scattering cross section times number density of photons in the field) times the average energy of the photons. 

Since the power in eq.~(\ref{PowerIC}) 
is a relativistic invariant, to get the expression in the Lab frame (where the electron is moving at $\beta$) it is enough to express the energy density $\tilde{u}$ in a manifestly invariant form. Remembering that $[u]=[\epsilon \times n]$, and that the number density $n$ transforms as a time since both $\d t \d^3x $  (four-volume) and $n \d^3x$ (number of particles) are relativistic invariants, we find that $u$ transforms as the square of the energy,
\begin{equation}
\tilde{u}=u\gamma^2(1-\beta\cos\tilde\theta)^2 \Longrightarrow \langle \tilde{u}\rangle=u\gamma^2\left(1+\frac{\beta^2}{3}\right)\,,
\end{equation}
where the last step holds for an isotropic radiation field. 
The energy lost by the electrons per unit time is the difference of the scattered power minus incoming power  $\sigma_T\,u$, hence 
\begin{equation}
-\frac{\d E}{\d t}=\sigma_T\,u\left[\gamma^2\left(1+\frac{\beta^2}{3}\right)-1\right]=\frac{4}{3}\gamma^2\beta^2\,u\sigma_T\simeq \frac{4}{3}\gamma^2\,u\sigma_T\,.\label{dedt}
\end{equation}

\subsection{Synchrotron radiation}
If $u$ is interpreted as the energy density of both photons and magnetic fields, eq.~(\ref{dedt}) describes not only inverse-Compton losses (in the Thomson regime), but also {\it synchrotron losses}, i.e. the power radiated when evolving in a magnetic field. This is due to the similar fundamental nature of both processes, see the right diagram of Fig.~\ref{fig:BasicQED}, describing synchrotron radiation as a scattering onto a virtual photon associated to an external magnetic field $B$~\footnote{Also bremsstrahlung can be considered Compton scattering of the CR electrons off the virtual photons of the {\it electrostatic} fields of the background charges: The right diagram of Fig.~\ref{fig:BasicQED} is formally identical to Fig.~\ref{fig:brems} when neglecting the recoil of the target. As detailed in~\cite{Blumenthal:1970gc}, the differential cross section $\d \sigma $ to emit a photon within $\d k$---in the so-called Weizs\"acker-Williams approximation valid for soft photons---is  $\d \sigma \simeq \sigma_T\d w$, where  $\d w\simeq \alpha \d k/k$ is the probability to emit a photon within $\d k$, as seen by inspecting eq.~(\ref{sigbrems}). For a pedagogical introduction, see~\cite{WW00}.}. To convince oneself of that, one can compute the power radiated according to Larmor's formula by a charge $q$ moving with Lorentz factor $\gamma$ with respect to the frame where a the static magnetic field is present.
In the CR  frame, the induced electric field and associated acceleration write
\begin{equation}
{\bf E}' = -\gamma{\bf v}\times {\bf B}\,\Longrightarrow {\bf a} = -\frac{q}{m}\gamma{\bf v}\times {\bf B}\,,
\end{equation}
hence, denoting $\zeta$ the angle of the velocity with the magnetic field, the radiated synchrotron power is 
\begin{equation}
P_s=\frac{2q^4\gamma^2}{3\,m^2}v^2B^2\sin^2\zeta\Rightarrow \langle P_s\rangle = \frac{4}{3}\sigma_T \gamma^2 u_B^2\,, \label{syncpower}
\end{equation}
with the angle-averaged expressed (assuming isotropy) is equivalent to eq.~(\ref{dedt}), once using $u=B^2/(8\pi)$!
For the equality to hold, it means that the frequency of the virtual photons associated to $B$ should verify the Thomson approximation. In order to check that, we now estimate the energy of synchrotron photons emitted in the Lab frame, verifying that they are indeed much smaller than the electron energy. 

As a preliminary result, we remind the concept of {\it beaming for relativistic motion}.
The time dilation and space contraction by the factor $\gamma$ in a (primed) frame moving at velocity $\beta$ with respect to the Lab one write 
\begin{equation}
t'=\gamma(t-\beta x)\,,\:\:\:x'=\gamma(x-\beta t)\,.
\end{equation}
From these transformations one derives
\begin{equation}
\frac{{\rm d}x}{{\rm d}t}=\frac{\gamma ({\rm d}x'+\beta {\rm d}t')}{\gamma ({\rm d}t'+\beta {\rm d}x')}=\frac{u_x'+\beta}{1+\beta u_x'}\,,
\end{equation}
and 
\begin{equation}
\frac{{\rm d}y}{{\rm d}t}=\frac{{\rm d}y'}{\gamma ({\rm d}t'+\beta {\rm d}x')}=\frac{u_y'}{\gamma(1+\beta u_x')}\,,
\end{equation}
where we introduced the velocity components $u_x'=\d x'/\d t'$, $u_y=\d y' / \d t'$.
This implies 
the aberration formula 
\begin{equation}
\tan\theta\equiv\frac{u_y}{u_x}=\frac{u_y'}{\gamma(u_x'+\beta )}=
\frac{\sin\theta'}{[\gamma(\beta +\cos\theta')]}\,.
\end{equation}

\begin{figure}[t]
\begin{center}
\includegraphics[scale=0.4]{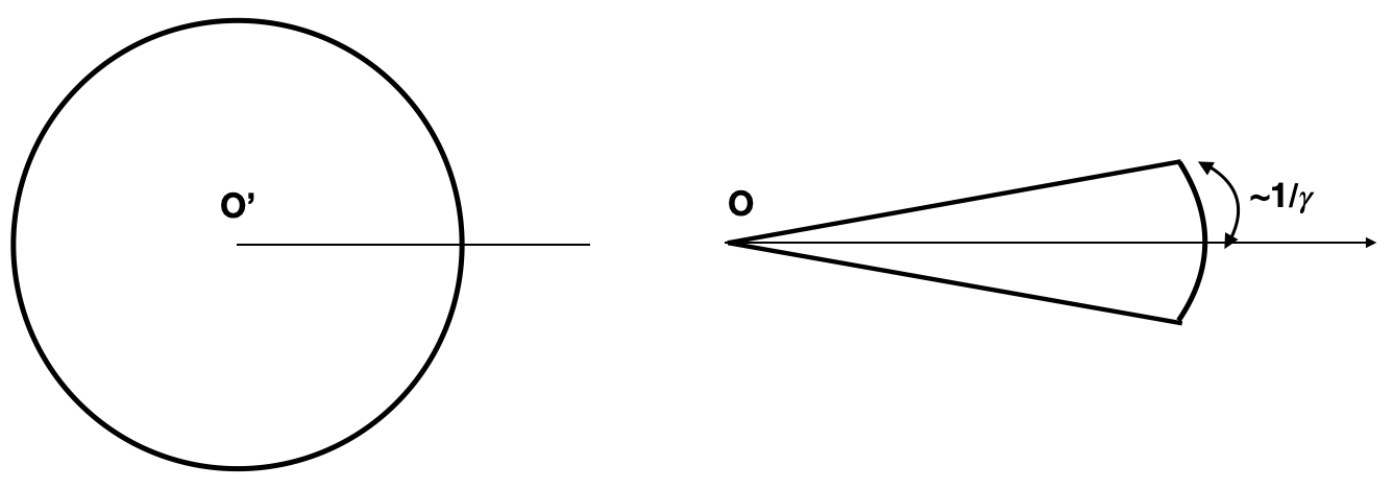}
\caption{Illustration of the beaming effect in the lab (unprimed frame) for an isotropic emission in the source (primed). 
\label{fig:beaming}}
\end{center}
\end{figure}

This means that a photon emitted at $\theta'=0$ travels at $\theta=0$, while a photon emitted  at $\theta'=\pi/2$ travels at $\sin\theta\simeq \theta\simeq1/\gamma$, as illustrated in Fig.~\ref{fig:beaming}.

\begin{figure}[t]
\begin{center}
\includegraphics[scale=0.45]{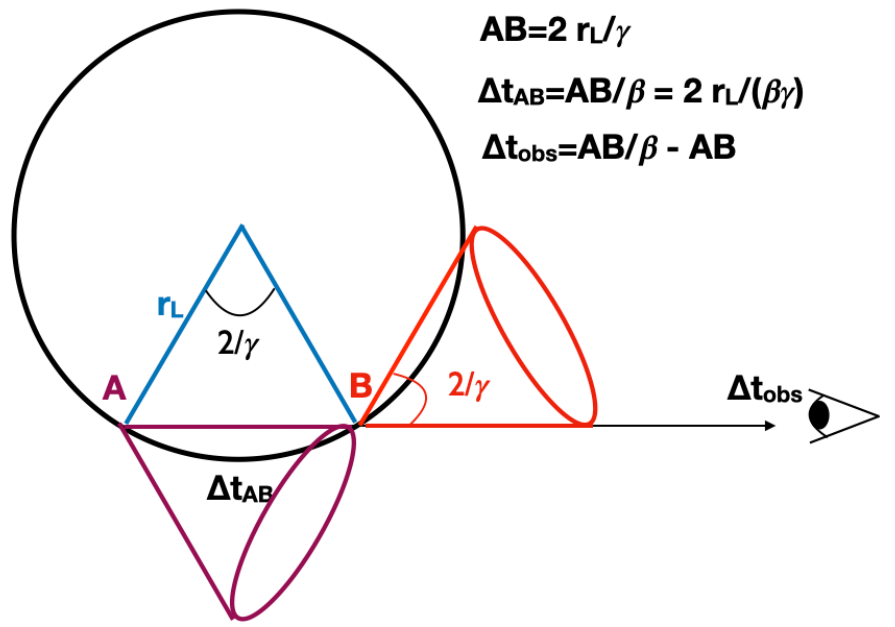}
\caption{Diagram to estimate the frequency of the synchrotron radiation, accounting for the beaming.\label{fig:beaming2}}
\end{center}
\end{figure}

To compute the spectrum of photons emitted in a time-varying emission process, one takes the Fourier transform of the signal (if this sounds unfamiliar, see e.g. chapter 2 in~\cite{RLrpap91}). In the non-relativistic limit, the circular uniform motion implies emission at the single frequency  (cyclotron line) $\nu_g=\frac{\omega_g}{2\pi}\equiv \frac{q\,B}{2\pi m}=2.8\,{\rm Hz}\,Z\,\left(\frac{m_e}{m}\right)\left(\frac{B}{\mu{\rm G}}\right)$. In the relativistic case, one must take into account that: i) the emission is observed only during the fraction of the orbit within the beaming angle; ii) the timescale of the observer is further different from the timescale of the emission; iii) the {\it gyroradius} $r_g=v_\perp/\omega_g$ is replaced by the Larmor radius $r_L=\gamma r_g$. As shown in Fig.~\ref{fig:beaming2},  
the time difference between passage at point B and point A is  given by
\begin{equation}
\Delta t\equiv t_B- t_A=\frac{AB}{\beta}=\frac{2\,r_L}{\gamma\beta}\,.
\end{equation}
 
For an observer  well beyond B along the direction AB,  the signal duration is
\begin{equation}
\Delta t_{\rm obs}=t_B+\delta t_{B}^{\rm prop}-(t_A+\delta t_{A}^{\rm prop})=
\frac{AB}{\beta}\left(1-\beta\right)=\frac{AB}{\beta}\frac{(1-\beta^2)}{1+\beta}\simeq \frac{AB}{\beta}\frac{1}{2\,\gamma^2}\simeq \frac{r_L}{\gamma^3}\,.
\end{equation}
As a poor man's proxy for a Fourier transform, we can estimate the  typical frequency of this {\it synchrotron radiation} as the inverse of the above, hence
\begin{equation}
\nu_s \simeq \frac{\gamma^3}{r_L}=\gamma^2\frac{\omega_g}{2\pi} \Rightarrow E_s\simeq 500\, \mu {\rm eV} \frac{B}{\mu G}\left(\frac{E_e}{\rm GeV}\right)^2\,.\label{nusEe}
\end{equation}
One can check that indeed the Thomson limit is well-respected for all energies of interest. 
A precise calculation could be based on the  Compton-scattered spectrum of the virtual photons of the $B-$field, obtained by going to the electron's rest frame. This approach  is however made cumbersome due to the need to transform into a {\it non-inertial} rest-frame. Hence, the spectrum is usually computed via the classical Li\'enard-Wiechert potentials seen by a distant observer, as  sketched in Appendix~\ref{synchdet}, based on chapter 3 of~\cite{RLrpap91}.  Such a calculation shows that  the power emitted per unit frequency scales as $\sim \nu^{1/3}$ at low frequencies and  as $\nu^{1/2}e^{-\nu/\nu_c}$ at high frequencies, with the critical frequency $\nu_c$ of the same order as the $\nu_s$ in eq.~(\ref{nusEe}), which does provide a reasonable proxy for the actual result.

Based on eq.~(\ref{syncpower}), the {\it power} emitted from a power-law $e$ spectrum  $\phi\propto E_e^{-\alpha}$ is

\begin{eqnarray}
&&\nonumber P_s\propto \int {\rm d}E_eB^2E_e^2 \delta(\nu_s-\kappa B\,E_e^2) E_e^{-\alpha} \\
&&\Longrightarrow \frac{\d P_s}{ {\rm d}\nu_s}
\propto\left.\frac{B^2E_e^{2-\alpha}}{{\rm d} \nu_s/{\rm d} E_e}\right|_{E_e(\nu_s)}\propto B^{\frac{1+\alpha}{2}}  \nu_s^{\frac{1-\alpha}{2}}\,.
\end{eqnarray}
Hence, from the slope of the synchrotron radiation power ($(1-\alpha)/2$) one can infer the slope of the parent electron distribution ($-\alpha$). This is a very useful tool to infer remotely the properties of non-thermal electrons at the sources, as opposed to the time and space integrated flux of CR electrons measured at Earth. Note that the same dependence holds for the the IC spectrum from a monochromatic photon background, if the scattering happens in the Thomson regime. Also pay attention to the (usually more than linear) dependence of the emitted power on the $B$-field strength: In cases where non-thermal emissions at shocks could be imaged in the X-rays, this has  been used to argue for magnetic field amplifications in the so-called rims at shocks, see for instance~\cite{Ressler:2014gza}.

\section{Hadronic interactions}\label{hadlosses}
\begin{figure}[t] 
\begin{center}
\includegraphics[scale=0.6]{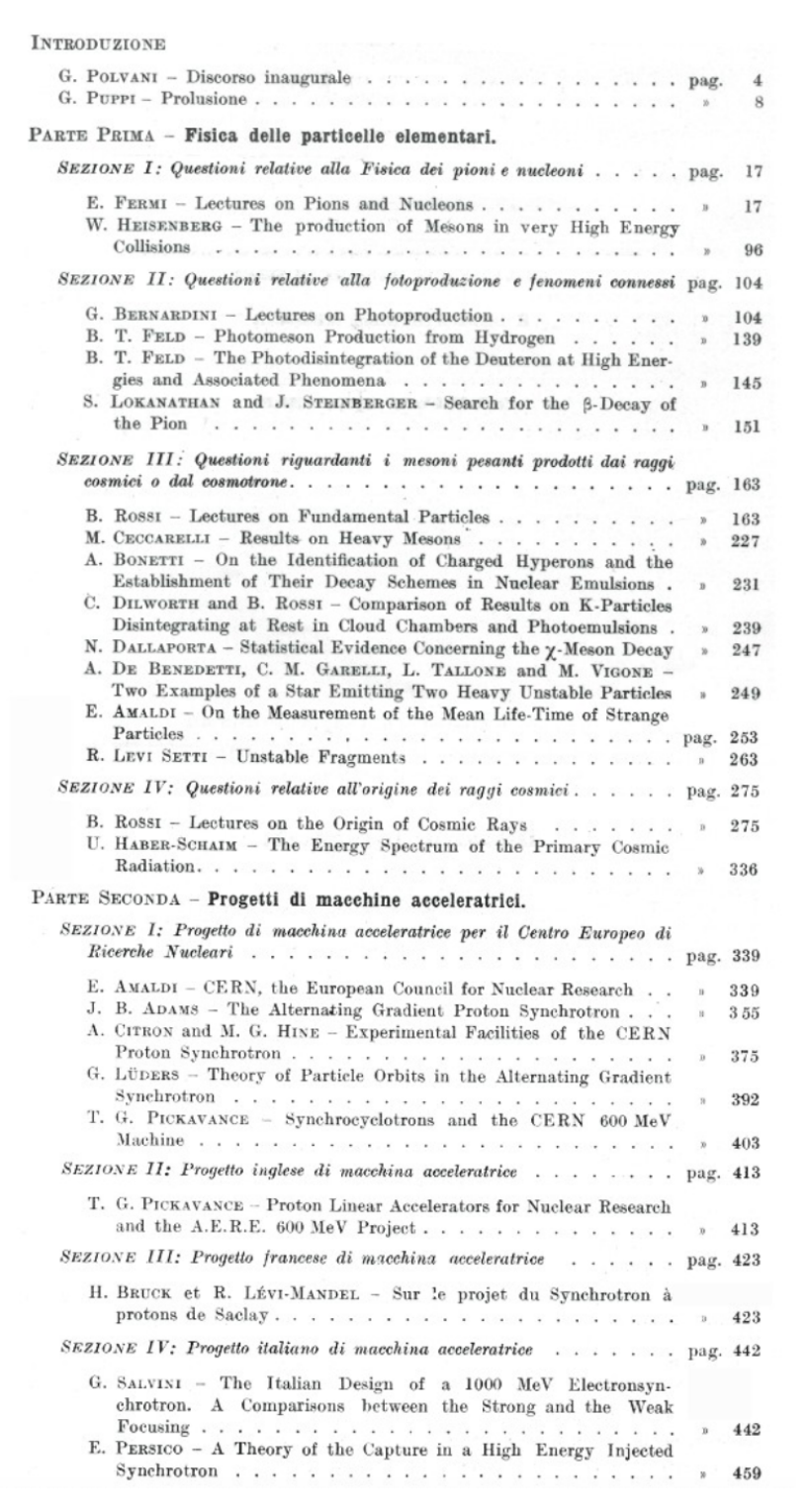}
\caption{The program of the second edition of the Varenna course, in 1954~\cite{Varenna54}~\label{program54}.}
\end{center}
\end{figure}

In principle, hadronic CR can lose energy  both via elastic and inelastic~\footnote{By inelastic we mean collisions where new particles are produced or the internal structure of the target nucleus or the projectile is changed. } collisions with target nuclei.  In practice, elastic scattering is usually disregarded: Given the CR kinematics, elastic cross-sections peak in the forward direction and the CR energy loss is negligible, see e.g. the comment in~\cite{Bergstrom:1999jc}.
{\it Inelastic losses} are however the dominant $E$-loss mechanism for nuclear kinetic energies above a few hundreds MeV/nuc, and will be our focus in what follows.

Major particle physics discoveries in the cosmic rays between $\sim 1930$ and mid-1950's (positron, muon, pion, strange hadrons) were due to inelastic hadronic interactions of CRs in the atmosphere. This was the central topic at the first gathering of this series in Varenna, in 1953, with lectures of Patrick M. S. Blackett and Cecil Powell among others~\cite{Varenna53}. It is also worth stressing how, the same year, the International Cosmic Ray Conference taking place at Bagn\`eres de Bigorre, France, marked a split between research in CR astrophysics and fundamental particle interactions, henceforth investigated more effectively at colliders (for an account of that conference, see e.g.~\cite{Cronin:2011zz}). This ``shift'' was clearly in the spirit of the time: The second edition of the Varenna course~\cite{Varenna54}, the 1954 one whose scientific program was inaugurated by Fermi a few months before his death, already saw a prominence of the accelerator approach to particle physics. The CR aspects were relegated to  Session IV and to some extent Session III of Part I, in which the dominant role was played by B. Rossi (see Fig.~\ref{program54}).

\begin{figure}[ht!]
\begin{center}
\includegraphics[scale=0.6]{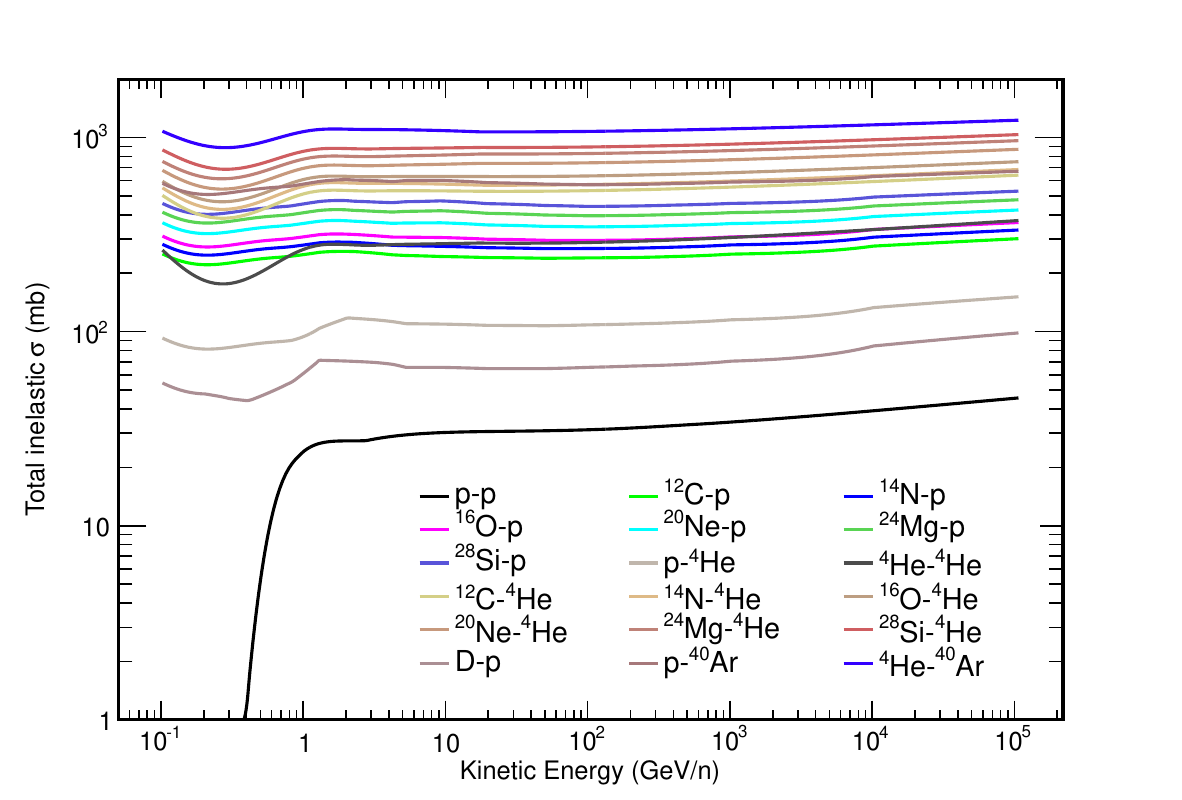}
\caption{Total inelastic cross sections as a function of the energy per nucleon of the incoming projectile.
The plot shows the cross sections for all the projectile-target pairs studied in~\cite{Mazziotta:2015uba}. Reproduced with the permission of the authors.}
\label{FigXsecInel}
\end{center}
\end{figure}

The current fundamental theory for strong interactions, QCD among quarks and gluons, is only perturbatively applicable to processes involving large exchanged momenta~\footnote{That is, well above $\Lambda_{\rm QCD}\sim 300\,$MeV, the {\it transmutation} energy scale below which,  loosely speaking, the {\it running} coupling constant of the QCD gauge theory explodes above unity.}, such as those studied at the LHC.  In the processes of interests to us, quarks and gluons within nucleons are not resolved; QCD is in its strongly coupled regime, and perturbation theory is inapplicable. A more useful way to think about nuclear interactions is via the effective Yukawa theory, involving nucleons and nuclei exchanging rather strongly coupled massive mediators (the pions), responsible for the intensity and relatively short range of the strong nuclear interaction. Since nuclear sizes are comparable or larger than the range of the Yukawa force, cross sections (dominated by the inelastic channels above  GeV/nucleon) scale as the geometric cross-section of the nuclei. Nuclei of mass number $A$ have a radius $R\simeq 1.2\,A^{1/3}\,$fm (with the $\sim$fm size of the nucleon essentially linked to $\Lambda_{\rm QCD}^{-1}$), hence the inelastic cross sections of a nucleus $\eta$ with a target $t$, $ \sigma_{\eta t}^{\rm inel}$, is expected to be $\sigma_{\eta t}^{\rm inel}\sim \pi (R_\eta^2+R_t^2)\sim 45\,(A_\eta^{2/3}+A_t^{2/3})\,{\rm mb}$, indeed close to proton-nuclear data above 2 GeV~\cite{Letaw:1983}.  Nuclear inelastic cross sections are weakly energy-dependent in the GeV-TeV range of primary interest here: For an illustration, see Fig.~\ref{FigXsecInel}, reporting a number of inelastic cross sections vs. kinetic energy/nucleon implemented in the \texttt{FLUKA} code~\footnote{http://www.fluka.org/fluka.php}. At lower energies, some energy dependence is present, for instance related to thresholds,  or manifesting a shallow minimum at few hundreds MeV/nuc and a rise at tens of MeV/nuc. In general, in this range nucleus-specific details may matter. 

An important energy benchmark is the threshold for pion production: The process $pp\to pp\pi^0$  has a kinetic energy threshold $K_{\pi^0}$
\begin{equation}
2m_{p}^{2}+2 (K_{p}+m_p) m_{p}>(2m_p+m_\pi)^2\Longrightarrow  K_{p}>K_{\pi^0}=2 m_\pi +  \frac{m_\pi^2}{2\,m_p}\simeq 290\,{\rm MeV}\,.\label{Thpion}
\end{equation}
The associated  $E$-loss term for a CR of mass number $A$ due to pion production can be approximated as~\cite{Krakau:2015bea}
\begin{equation}
-\left(\frac{\d E}{\d t}\right)=3.85 \times 10^{-16}A^{0.79}\left(\frac{n_{\text {gas }}}{\mathrm{cm}^{-3}}\right)\left(\frac{E}{\mathrm{GeV}}\right)^{1.28} \times\left(\frac{E}{\mathrm{GeV}}+200\right)^{-0.2} \mathrm{GeV} \mathrm{s}^{-1}\,,
\end{equation}
according to the re-assessement of modern parameterization of pion production in~\cite{Kelner:2006tc}.

Below the energy (per nucleon) given in eq.~(\ref{Thpion}), the only inelastic processes that matter are {\it spallation} ones, where target and/or projectile nuclei are broken into multiple nucleon/nuclear fragments. 
Note that these processes are the leading source of 
nuclei such as Li-Be-B, which are {\it fragile} from the thermonuclear point of view: Due to their small nuclear binding energy, they are easily burned in stellar thermonuclear processes and are only present in traces in conventional astrophysical environments, such as the ISM. Their sizeable presence among CR nuclear fluxes is interpreted as the consequence of spallation of heavier CRs common in the ISM, such as C or O, onto the ISM gas during their propagation. Such nuclei are thus also referred to as ``secondary'' CR species. Since single-nucleons are rather common among fragments, spallation processes also contribute as a non-negligible secondary source term for protons.

 Due to theoretical limitations, one must rely on semi-empirical formulae  to describe not only the inelastic cross sections,   but also the relevant {\it exclusive} differential cross-sections to produce the secondary $\alpha$ in the collision of the primary $\eta$ with the target $t$. They can be written as function of kinetic energy of the primary, $K_\eta$, and product, $K$, as
  \begin{equation}
\frac{\d \sigma_{\eta+t\to\alpha}}{\d K}(K,K_\eta) = \sigma_{\eta t}^{\rm inel}(K_\eta)\frac{\d {\cal N}_{\eta+t\to\alpha}}{\d K}(K,K_\eta)\,,\label{generalparam}
\end{equation}
with $\d {\cal N}/\d K$ being the multiplicity spectrum of the species $\alpha$ in the collision of $\eta$ with $t$.  Since data are typically scarce, regularities motivating  semi-empirical formulae turn out to be useful in interpolating between and extrapolating beyond measurements, or to estimate cross-sections involving nuclei for which no measurement exists. Calibrated models and codes for the nuclear cross sections libraries  are implemented and described in propagation codes, such~\cite{Strong:1998pw,Evoli:2017vim,Genolini:2018ekk}. Some relevant cross-sections and branching ratios are also reported in  printed form in Tab. 10.1 of~\cite{Longair}.

\subsection{Examples: Spallations, hadronic gamma-rays, antiprotons}
As an example of some of these regularities,  well away from thresholds, $\d {\cal N}/\d K$ is prominently a function of $x\equiv K/K_\eta$ but is only weakly dependent from $K$, and depends on $\eta$ and $t$ mostly via a normalisation (equivalent to the branching ratio $\kappa$). To a good approximation, estimated to a 5-6\% level for B/C in~\cite{Maurin:2001yxa}, in spallation reactions the kinetic energy per nucleon is conserved, so that
  \begin{equation}
\frac{\d {\cal N}_{\eta+t\to\alpha}}{\d K}(K,K_\eta)\simeq \kappa_{\eta+t\to\alpha}\delta\left(\frac{K}{A_\alpha}-\frac{K_\eta}{A_\eta}\right)\,.
\end{equation}

It is worth mentioning that the projectile or target nucleus can end up in an unstable state and de-excite via gamma-ray emission, at typical energies of few MeV for nuclei at rest. Although this has long been recognized as a potential exquisite diagnostic tool for the study of energetic phenomena~\cite{Ramaty:1979}, the observational challenges in MeV gamma-ray astronomy make this field still in its infancy, at least for CRs in the ISM. 

While inelastic processes must be taken into account for precision calculations of hadronic CR spectra,  at sufficiently high energy (typically above tens of GeV/nuc) they have a sub-leading role for primaries compared to collisionless propagation effects, as it will be illustrated in a simplified propagation model in Sec.~\ref{soverp}. Nonetheless, at all energies they are 
 of crucial importance as {\it source terms} not only for secondary CR nuclei, but also for hadronic gamma-rays, neutrinos, and {\it antinuclei}. 

As a specific example of this link, let us focus on the hadronic photons, i.e. essentially those produced via the neutral pion decay process, $\pi^0\to \gamma\gamma$. 
The photons are back-to-back in the $\pi^0$ frame, each carrying  an energy equal to $m_\pi/2\simeq 67.5\,$MeV. If the pion moves at $\beta$, in the Lab frame the energy of the photon  is
$E_\gamma=m_\pi\gamma(1+\beta \cos\theta)/2$, $\theta$ being the angle of the emitted photons with respect to the the direction of flight of the pion. Hence, the maximal and minimal energy of the photons obey
\begin{equation}
E^{\rm max}_{\rm min}=\frac{m_\pi}{2}\gamma (1\pm \beta)\,\Longrightarrow E_\gamma^{\rm max}E_\gamma^{\rm min}=\frac{m_\pi^2}{4}\,,\:\:\:E_\gamma^{\rm max}+E_\gamma^{\rm min}=E_\pi\,.
\end{equation}
There is a one-to-one correspondence between $E_\gamma$ and $\theta$, whose law is
\begin{equation}
{\rm d}E =\frac{m_\pi}{2}\gamma\beta{\rm d}\cos\theta\,.
\end{equation}

Since the pion is a scalar particle, its decay products are emitted isotropically, hence we can easily deduce the energy distribution of the photons: 
\begin{equation}
\frac{{\rm d}N}{{\rm d}\Omega}=\frac{1}{4\pi}\Longrightarrow{\rm d}N =\frac{1}{2}{\rm d}\cos\theta\Longrightarrow  \frac{{\rm d}N}{{\rm d}E}=
\frac{1}{m_\pi\gamma\beta}=\frac{1}{E_\pi\beta}=\frac{1}{\sqrt{E_\pi^2-m_\pi^2}}\,.\label{angdistrphot}
\end{equation}
This distribution is flat (box-shaped) in energy space between $E_{\rm min}\simeq 0$ and $E^{\rm max}\simeq E_\pi$ in the relativistic limit. In log-energy space,
\begin{equation}
\frac{1}{2}(\log E^{\rm min}+\log E^{\rm max})=\log \sqrt{E^{\rm min}E^{\rm max}}=\log \left(\frac{m_\pi}{2}\right)\,,\label{Epeak}
\end{equation}
i.e. {\it the center of the interval is half the pion mass, independently of the pion energy distribution, hence of the parent nucleon distribution}. Adding an arbitrary number of box-shaped spectra,
each centred at $m_\pi/2$, implies that this is the maximum of the spectrum. This is dubbed ``pion bump'' and considered the cleanest (albeit hard-to-detect!) signature of hadronic origin of a gamma-ray spectrum~\cite{AGILE:2011tzq,Fermi-LAT:2013iui}.

Let us assume for simplicity a pure hydrogen target of density $n$  and relativistic projectiles and products so that $K\simeq E$.
A useful approximation for the spectrum of pions from a proton CR interaction is 
\begin{equation}
\frac{\d {\cal N}_{p+H\to\pi}}{\d E}(E,E_p)\simeq \zeta_\pi\delta(E- \kappa_\pi E_p)\,,
\end{equation}
where the pion multiplicity $\zeta_\pi$ and the average fraction of the proton energy into a pion, $\kappa_\pi$, are weakly dependent functions of energy, neglected below.  The pion source term writes 
\begin{equation}
q_\pi (E)=n\int {\rm d}E_p\frac{\d {\cal N}_{p+H\to\pi}}{\d E}\sigma_{pp}^{\rm inel}(E_p) \phi_p( E_p)=\frac{n\,\zeta_\pi}{\kappa_\pi}\,\sigma_{pp}^{\rm inel}\left(\frac{E}{\kappa_\pi}\right) \phi_p\left(\frac{E}{\kappa_\pi}\right)\,,
\end{equation}
where $\sigma_{pp}^{\rm inel}(E_p)$ is the proton-proton inelastic cross section. For more advanced treatments, see e.g.~\cite{Kelner:2006tc} or~\cite{Kafexhiu:2014cua}.

To a first approximation, the pion source spectrum at $E_\pi$ is thus proportional to the proton flux at an energy $E_p=E_\pi/\kappa_\pi$ (with sizable corrections due to the cross section energy dependence, relevant close to threshold). The photon spectrum is then
\begin{equation}
q_\gamma (E_\gamma)=2\int_{E_{\pi}^{\rm min}(E_\gamma)}^\infty {\rm d}E_\pi\frac{\d N}{\d E}q(E_\pi)=2\int_{E_\gamma+\frac{m_\pi^2}{4E_\gamma}}^\infty {\rm d}E_\pi\frac{q(E_\pi)}{\sqrt{E_\pi^2-m_\pi^2}}\,,\label{hadgsource}
\end{equation}
where the minimum energy of the pion to produce a photon of energy $E_\gamma$, $E_{\pi}^{\rm min}(E_\gamma)$, is given by the relation linking the maximal energy of a photon produced by a pion of $E_\pi$. 
For a given energy $E_\pi$, $E_\gamma^{\rm max}$ and conversely $E_{\pi}^{\rm min}(E_\gamma)$ are given by
\begin{equation}
E_\gamma^{\rm max}=E_\pi-E_\gamma^{\rm min}=E_\pi-\frac{m_\pi^2}{4E_\gamma^{\rm max}}\,,\Longrightarrow E_{\pi}^{\rm min}(E_\gamma)=E_\gamma+\frac{m_\pi^2}{4E_\gamma}\,.
\end{equation}

\begin{figure}[hb!]
\begin{center}
\includegraphics[scale=0.5]{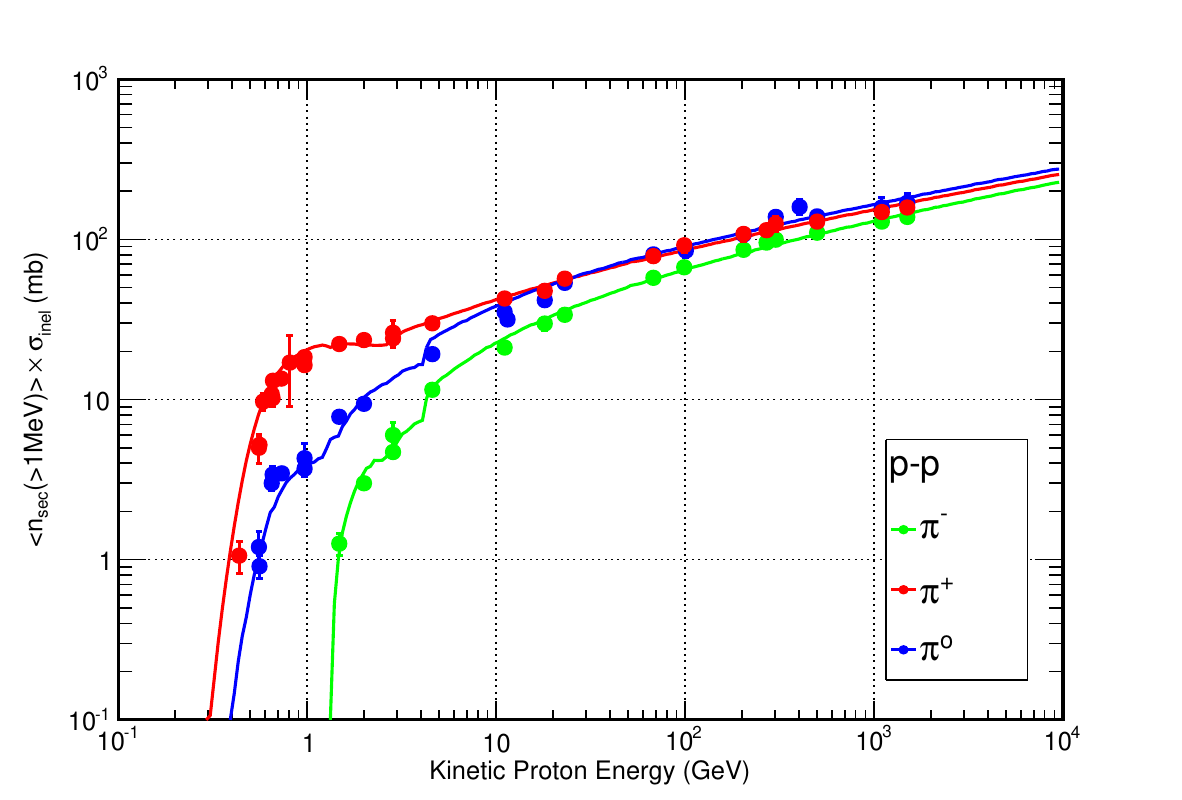}
\caption{Inclusive cross sections for the production of $\pi^0$ (blue), $\pi^{+}$ (red) and $\pi^{-}$ (green)
in $p-p$ collision as function of the incoming proton kinetic energy. 
Lines: {\tt FLUKA} simulation; points: data from Ref.~\cite{dermer1986}. From~\cite{Mazziotta:2015uba}, reproduced with the permission of the authors.}
\label{FigXsecPion}
\end{center}
\end{figure}

It is worth noting that each process producing a $\pi^0$ is associated to a process producing charged pions: For the $pp$ process, at CR energies one is often well above threshold, multipion production is frequent, and isospin symmetry yields equal numbers of $\pi^0$, $\pi^+$, and $\pi^-$, see Fig.~\ref{FigXsecPion}. 
Also, the energies of the gammas and the neutrinos are  within a factor of $\sim$two of each other: each of the there neutrinos emitted in a charged pion plus following muon decay carries about 1/4 of the energy of the parent pion, while photons exactly a fraction 1/2 of it. We thus expect a link between neutrino and (hadronic) photon spectra at emission given by 
\begin{equation}\label{eq:nugamma}
\frac{1}{2}\frac{{\rm d}N_\gamma}{{\rm d}E_\gamma} \simeq \frac{1}{6} \left[\, \frac{{\rm d}N_\nu}{{\rm d}E_\nu}\right]_{E_\nu = E_\gamma/2}~,
\end{equation}
which has been (and still is) extremely useful to estimate neutrino target fluxes as counterparts of observed photon sources (assumed hadronic) as well as in constraining interpretations of the  diffuse (probably extragalactic) neutrino flux observed by IceCube (for a review, see~\cite{Halzen:2021ynx}) based on gamma observations. Further details on neutrino fluxes can be found in Denise Boncioli's lectures.

Another important class of secondaries are antinucleons. Currently, only antiprotons have been measured (rather precisely) in CRs, and their spectrum is consistent within errors with expectations from secondary production~\cite{Boudaud:2019efq}. Their production cross-sections cannot be computed from first principle either, and are typically parameterised as
\begin{equation}\label{eq:xantip}
\frac{{\rm d} \sigma}{{\rm d} K_{\bar{p}}}=2 \pi p_{\bar{p}}(K_{\bar{p}}) \int_{\eta_{\rm min}\simeq 0}^{\infty} \d \eta \frac{1}{\cosh ^{2}(\eta)} \sigma_{\mathrm{inv}}\,,
\end{equation}
where $\eta=-\ln[\tan(\theta/2)]$ is the pseudo-rapidity, defined in terms of the scattering angle $\theta$, and
\begin{equation}\label{eq:siginv}
\sigma_{\mathrm{inv}} \equiv E \frac{\d^{3} \sigma}{\d p^{3}}\left(\sqrt{s}, x_{\mathrm{R}}, p_{\mathrm{T}}\right)
\end{equation}
depends only on manifestly Lorentz invariant quantities: the CoM energy $\sqrt{s}$, the transverse momentum of the produced antiproton ($p_T$) and the ratio of the antiproton energy to the maximally possible energy in the CoM frame ($x_R$). Since in all cases known to the author one factorises $ \sigma_{\mathrm{inv}} =\sigma_{pp}^{\rm inel}(s)\times F(s, p_T, x_R) $, and (in the case of proton-proton collisions) $s=2m_p(E_p+m)$, eq.~(\ref{eq:xantip}) is ultimately equivalent to a rewriting of eq.~(\ref{generalparam}). Several parameterisations have been proposed for antiproton cross-sections in the past decade, benchmarked to a growing dataset of collider data~\cite{diMauro:2014zea,Lin:2016ezz,Winkler:2017xor,Donato:2017ywo,Korsmeier:2018gcy}, with current uncertainties still among the dominant contributions to the total error budget of secondary CR antiproton fluxes~\cite{Boudaud:2019efq}.  Additional cross-sections inputs concern the antiproton energy-losses: Customarily, the ``sink'' term includes the total inelastic antiproton cross-section. Then,  a so-called tertiary source term enters, which is due to inelastically scattered secondary antiprotons. Its differential cross section is typically parameterised as
\begin{equation}
\frac{\d \sigma_{\bar{p}+p\to\bar{p}}}{\d K}(K,K_{\bar{p}}) = \frac{\sigma^{\rm inel}_{\bar{p}+p}-\sigma^{\rm ann}_{\bar{p}+p}}{K}\,.
\end{equation}
References containing fits to the relevant cross sections can be found e.g. in~\cite{Evoli:2017vim}.

Antiprotons are an important diagnostic tool especially for indirect searches aimed at identifying the nature of the still mysterious dark matter pervading the universe (for a recent study and assessment of the literature, see~\cite{Calore:2022stf}); hopefully these searches will be extended in the near future to light antinuclei such as antideuteron~\cite{Donato:1999gy} and antihelium~\cite{Cirelli:2014qia,Carlson:2014ssa}, with peculiar kinematical aspects leading to spectra features advantageous for exotic searches. For instance, antideuteron production in secondary reactions requires an energy of at least 17 $m_p$ (just apply the consideration developed after eq.~(\ref{sinvariant}) to the process $pp\to p\bar{p}n\bar{n}pp$). This makes an antideuteron low-energy component (which does not suffer from the Lab frame kinematical suppression above) an exquisite target for CR-based searches of DM, notably with the innovative GAPS experiment based on X-ray de-excitation of exotic atoms~\cite{Mori:2001dv}. Important uncertainties remain in modeling the production of CR antinuclei both in conventional and exotic processes, an area of fervent activity today (see e.g.~\cite{Serksnyte:2022onw,Shukla:2020bql}) which we cannot make justice to here.

\section{Photon interactions}\label{PhInter}
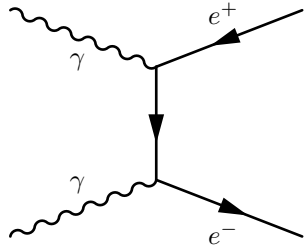
\begin{figure}
\begin{center}
    \begin{fmffile}{fgammagamma}
      \setlength{\unitlength}{0.6cm}
      \begin{fmfgraph*}(8,5)
        \fmfleft{i1,i0}
        \fmfright{o1,o0}
        \fmf{boson,label=$\gamma$}{i1,w1}
        \fmf{fermion,label=$e^-$}{w1,o1}
        \fmf{fermion}{w0,w1}
        \fmf{boson,label=$\gamma$}{i0,w0}
        \fmf{fermion,label=$e^+$}{o0,w0}
      \end{fmfgraph*}
    \end{fmffile}
    \caption{Diagram for the QED pair production process.}\label{pair}
    \end{center}
\end{figure}      

We have seen how loss processes of charged particles have their counterpart as source processes for non-thermal photons. Synchrotron photons associated to GeV to TeV electrons typically fall in the radio to X-ray band (see eq.~(\ref{nusEe})), while inverse Compton photons are typically from the soft to very high-energy gamma band  (see eq.s~(\ref{Then}, \ref{KNen})). Similarly, photons from pion decay also fall in the gamma-ray band (see eq.~(\ref{hadgsource})). These photons are important messengers 
of their parent charged particles, since they retain directionality and can be associated way more easily to astrophysical sources.  It is important to realise, however, that photons as well are subject to scattering and absorption. This process due to the interstellar gas and especially dust (onto which photons can coherently diffuse via {\it Rayleigh scattering}) is typically known as {\it extinction} in low-energy astronomy and is perhaps familiar to the reader.  In the extreme UV to X-ray range, photo-ionisation processes are relevant, but neither of these regimes will be dealt with here. At  very high energies, interactions with interstellar and a fortiori intergalactic matter are negligible, but the interaction with the radiation backgrounds are increasingly important.
The most important process is {\it pair production}, see Fig.~\ref{pair}, by a photon with high energy $E$ impinging on a background one of energy $\epsilon$.  The threshold for the reaction $\gamma+\gamma\to e^+e^-$  in the Lab is 
\begin{equation}
4E_{\gamma\gamma}^{\rm th}\epsilon=(2m_e)^2 \Longrightarrow E_{\gamma\gamma}^{\rm th}= \frac{m_e^2}{\epsilon}\,.
\end{equation}
The cross-sections behaves as
\begin{equation}
\sigma_{\gamma \gamma}\left(\beta\right)=\frac{3 \pi \sigma_T}{16}\left(1-\beta^{2}\right)\left[2 \beta\left(\beta^{2}-2\right)+\left(3-\beta^{4}\right) \ln \left(\frac{1+\beta}{1-\beta}\right)\right]\,,
\end{equation}
where $\beta=\sqrt{1-4m_e^2/s}$ is either lepton velocity in the CoM frame; $\sigma_{\gamma \gamma}$ peaks at about $\sigma_T/4$ at about twice the threshold energy.
It is easy to check that the $e^\pm$ pair production mechanism constitutes a serious limitation to measure photons from  the remote extragalactic sky (beyond redshift $z\sim 1$) already at $\sim 100$ GeV, for the near extragalactic sky (beyond $z\sim 0.1$) at $\sim$10 TeV, and even for Galactic objects in the PeV range: Beyond the TeV range, we can only perform astronomy in the ``local neighborhood'', which is one of the motivations to develop  neutrino astronomy at higher energies.

An interesting feature due to collisional effects involving $\gamma$'s and $e^\pm$ is that extragalactic photons initiate {\it electromagnetic cascades}: After the first pair-production, typically on EBL photons, affecting photons with $E_{\gamma}>\mathcal{E}_{\gamma}\equiv \frac{m_e^2}{\epsilon_{\rm EBL}}\simeq 390\,{\rm GeV}$, the $e^\pm$ scatter via inverse Compton onto the CMB resulting into highly energetic photons (notably at early stages in the Klein-Nishina regime, see eq.~(\ref{KNen})), which undergo the same multiplicative process as long as  $E_{\gamma}>\mathcal{E}_{\gamma}$. Another characteristic energy  is $\mathcal{E}_{X} \equiv \frac{1}{3}\mathcal{E}_{\gamma}\frac{\epsilon_{\rm CMB}}{\epsilon_{\rm EBL}}\simeq 1.2\times 10^{8}\,{\rm eV}\,,$ corresponding to the upscattered photon energy (in the Thomson regime) associated to minimum-energy $e^\pm$ produced by photons at the threshold $\mathcal{E}_{\gamma}$. Below $\mathcal{E}_{X}$, the spectrum is solely due to the further energy losses  of ``cascade-sterile'' $e^\pm$ via inverse Compton. 

While the exact spectral shape of the diffuse $\gamma$-ray flux depends on the cosmic evolution and the distance to the sources, in the limit of fully developed cascades,  one can derive analytically an approximate, universal spectral shape for the resulting diffuse $\gamma$-ray flux~\cite{AstroCR,Berezinsky:2016feh}, given by:
\begin{equation}
\phi_{\gamma}\left(E_{\gamma}\right)=\frac{E_{s}}{\mathcal{E}_{X}^2\left(2+\ln \mathcal{E}_{\gamma} / \mathcal{E}_{X}\right)}\times\left\{\begin{array}{ll}
\left(E_{\gamma} / \mathcal{E}_{X}\right)^{-3 / 2} & {\rm at }\:\:\: E_{\gamma} \leq \mathcal{E}_{X} \\
\left(E_{\gamma} / \mathcal{E}_{X}\right)^{-2} &  {\rm at }\:\:\: \mathcal{E}_{X} \leq E_{\gamma} \leq \mathcal{E}_{\gamma} \\
0 &  {\rm at }\:\:\: E_{\gamma}>\mathcal{E}_{\gamma}
\end{array}\right.\label{univgamma}
\end{equation}
where the normalisation is given in terms of the total injection energy $E_s$. After all the scatterings, no directional memory remains of the initial photons, but the quasi-isotropic extragalactic gamma-ray flux below the TeV range puts an upper limit to {\it any extragalactic photon source at supra-TeV energies}. This has been used in countless astroparticle applications. For instance, given the expected link between the spectrum of neutrinos and the hadronic gamma-ray spectrum (both originating via pion decays)  {\it at the source},
this has been used in the literature to raise the possibility that most of the sources contributing to the IceCube flux are actually {\it opaque} (i.e. the associated gamma-rays are degraded to MeV or even thermal energies before escaping the sources). For a recent calculation in that sense, see for instance~\cite{Capanema:2020oet}.

\section{Applications to cosmic ray propagation}  \label{Applications} 
In the hybrid (non-inertial) plasma frame, where positions are measured in the Lab (Galaxy) frame but particle momenta are measured with respect to where magnetic inhomogeneities are statistically at rest, the ensemble average (over realisations of inhomogeneities) and angle average of the single particle distribution function $\phi_\alpha$ for a CR species $\alpha$ obeys (sum convention over repeated {\it latin} indices)
\begin{eqnarray}
&&\label{comppropeq} \frac{\partial \phi_\alpha}{\partial t}-\frac{\partial }{\partial x_i}D_{ij}\frac{\partial \phi_\alpha}{\partial x_j}+u_i\frac{\partial \phi_\alpha}{\partial x_i}-\frac{1}{3}\frac{\partial u_i}{\partial x_i}\left(p\frac{\partial \phi_\alpha}{\partial p}\right)-\frac{1}{p^2}\frac{\partial  }{\partial p}\left(p^2{\cal D}\frac{\partial \phi_\alpha}{\partial p}\right)=\\
&& q{\color{blue}+\frac{1}{p^2}\frac{\partial}{\partial p}\left[p^2 \left(\frac{\d p}{\d t}\right)_{\ell}\phi_\alpha\right]}{\color{red}-\Gamma_{\rm tot,\alpha}\,\phi_\alpha}{\color{magenta}+\sum_\eta \phi_{\eta}\otimes\Gamma_{\eta\to\alpha}}\,,\nonumber
\end{eqnarray}
where the LHS contains transport terms whose origin is collisionless (from second to fifth terms: spatial diffusion with diffusion tensor $D_{ij}$, convective term associated to the plasma velocity field $u_i({\bf x})$, adiabatic energy loss/gains, and reacceleration or magnetic inhomogeneity diffusion in momentum space, controlled by ${\cal D}$), while the RHS, besides a possible primary source term $q$, contains terms describing continuous losses (in blue) and catastrophic sinks (in red) or sources (in magenta).  Catastrophic sinks include both decays if the species is unstable with lifetime at rest $\tau_{\rm dec}$ (one should account for the boost $\gamma$ of the CR in the Galactic frame), and inelastic interactions over all possible targets $t$ of the ISM with density $n_t$, so that
  \begin{equation}
\Gamma_{\rm tot,\alpha}=\frac{1}{\gamma \tau_{\rm dec}^\alpha}+\sum_{ t} \beta \sigma_{\alpha t}^{\rm inel} n_t\,.
\end{equation}
In practice, only hydrogen and helium have significant densities to be relevant ISM targets.
The last term in eq.~(\ref{comppropeq}) represents in general an integral operator acting on $\phi_\beta$, which is more easily expressed in terms of the kinetic energy  $K=E-m$ rather than momentum variables~\footnote{Note that $\phi(K)=\phi(p(K))\d p/\d K=\beta^{-1}\phi(p(K))$, since $p^2=(m+K)^2-m^2=2mK+K^2$.}, so that
  \begin{equation}
\phi_{\eta}\otimes\Gamma_{\eta\to\alpha}=\sum_{ t} n_t \int \d K_\eta \beta \frac{\d \sigma_{\eta+t\to\alpha}}{\d K}(K,K_\eta) \phi_\eta(K_\eta)\,.\label{spallsource}
\end{equation}

Since each term in eq.~(\ref{comppropeq}) has the dimension of $\phi/{\rm time}$, one can define characteristic timescales (analogous to eq.~(\ref{tauloss})) which allow for a quick parametric assessment of their importance. Public codes exist that allow for a numerical (such as \texttt{GALPROP}~\footnote{\texttt{https://galprop.stanford.edu}} \texttt{DRAGON}~\footnote{\texttt{https://github.com/cosmicrays/}}) or semi-analytical (\texttt{USINE}~\footnote{\texttt{https://dmaurin.gitlab.io/USINE/}}) solution of the problem. It is important however to grasp the key features of these terms via some analytical, limiting solution, notably of the steady state problem. We outline two examples below, which have the merit to show the interplay of fundamental physics properties of the collisional processes with CR phenomenology.

\subsection{Continuous E-loss dominated propagation}
If continuous energy loss timescales are the shortest ones (or the only one of relevance, in ``quasi-homogeneous'' problems), the steady state equation approximates to
  \begin{equation}
-\frac{1}{p^2}\frac{\partial}{\partial p}\left[p^2\left(\frac{\d p}{\d t}\right)_{\ell}\phi_\alpha\right]=q\,,
\end{equation}
and is solved by
  \begin{equation}
\phi(p) \propto -\frac{1}{p^2(\d p/\d t)_{\ell}} \int^p {\rm d} p' q(p')\,p'^2\,,
\end{equation}
which, for $q\propto p^{-s}$ and  $(\d p/\d t)_{\ell}\propto -p^{\ell}$, leads to 
 \begin{equation}
\phi(p) \propto p^{-s-\ell+1}\,.\label{specelosses}
\end{equation}
Namely, {\it the steady state spectrum is $\ell-1$ softer than the injected one}. It turns out that for CR leptons the continuous $E$-loss dominance is close to truth, with ionisation and Coulomb losses ($\ell\simeq 0$) dominating at low-energies, eventually
overcome by bremsstrahlung energy losses ($\ell\simeq 1$) and Compton and Synchrotron energy losses ($\ell\simeq 2$) finally taking over. As a result, as a first approximation we expect the low-$E$ (sub-GeV) electron spectrum observed at the Earth to be one power harder than the source one, that the spectrum matches the source one at intermediate energies (few GeV), while being one power steeper than the injected one at high-energies. Of course, diffusion and other transport terms modify these basic features, which remain however qualitatively correct.

\subsection{Catastrophic loss for diagnostics: Secondaries over primaries}~\label{soverp}
Let us compare secondaries, i.e. nuclei only produced by spallation during propagation, such as the above-mentioned
 Boron, Lithium, and Beryllium, with primaries. For simplicity, let us assume that primary CR sources are confined to a thin disk of half-width $h$, practically infinitesimal if compared to the diffusive propagation halo half-thickness $H$.  We are then reduced to a 1D problem is considering the disk as infinite, translating the fact that the radial extent of the Galaxy is large compared to its vertical extension.  
 If we include just the dominant diffusive transport operator, assumed isotropic, the steady state transport equation for a purely {\it primary} species reduces to 
  \begin{equation}
-\frac{\partial}{\partial z}\left(D\frac{\partial \phi^P}{\partial z}\right)=2\,q_0(p)\,h\delta(z)-2h\,\Gamma_\sigma\,\phi^P\,\delta(z)\,\label{slab1P}\,.
\end{equation}
At $z\neq 0$, eq.~(\ref{slab1P}) reduces to $\frac{\partial^2 \phi^P}{\partial z^2}=0$, whose solution is $\phi^P(z,p)=a(p)+b(p)|z|$. If we denote with $\phi_0^P$ the solution
in the plane $\phi_0^P=\phi^P(0,p)$, the vanishing at the boundary $|z|=H$ gives
 \begin{equation}
\phi^P(z,p)=\phi_0^P(p)\left(1-\frac{|z|}{H}\right)\,.\label{fz}
\end{equation}
An equation for $\phi_0^P(p)$ can be found by integrating Eq.~(\ref{slab1P}) over a small interval around $z=0$, 
\begin{equation}
-2D(p)\left.\frac{\partial\,\phi^P}{\partial z}\right|_0=2h(q_0(p)-\Gamma_\sigma\phi_0^P)\,\label{slab2}\,.
\end{equation}
We are further assuming the diffusive halo  to be homogeneous, although $D$ can depend on $p$. 
Using Eq.~(\ref{fz}) one finds 
\begin{equation}
\phi_0^P(p)=q_0(p)\tau_{\rm eff}(p)\,,\:\:\:\:{\rm where}\:\:\:\:\tau_{\rm eff}^{-1}(p)=\tau_d^{-1}(p)+\tau_\sigma^{-1}(p)\,\label{slab3P}\,,
\end{equation}
where we defined a diffusive timescale $\tau_d$
\begin{equation}
\tau_d(p)\equiv\frac{H\,h}{D(p)}\approx10^7\,{\rm yr}\frac{H}{3\,{\rm kpc}}\frac{h}{100\,{\rm pc}}\frac{10^{28}\,{\rm cm}^2{\rm s}^{-1}}{D}\label{taud}\,,
\end{equation}
and a catastrophic loss timescale $\tau_\sigma$
\begin{equation}
\tau_{\rm \sigma}(p)\equiv \Gamma_{\sigma}^{-1}\approx10^7\,{\rm yr}\left(\frac{1\,{\rm cm}^{-3}}{n_{\rm ISM}}\right)\left(\frac{100\,{\rm mb}}{\sigma}\right) \label{tausig}\,.
\end{equation}
Note that, as long as $D(p)$ is a growing function of $p$, the diffusive timescale $\tau_{d}$ dominates over collisional losses at sufficiently high energies, since  $\Gamma_\sigma \simeq \sigma \,n $ has very little dependence on $p$, as we described. This remains true in more general models: {\it Collisional effects are more relevant in shaping hadronic CR fluxes at low energies}.

Secondaries  $\phi^S(p)$ are sourced by the spallated parents per unit time, i.e. $q_0(p)\to \phi_P\Gamma_{P\to S}$, with collision only possible in the thin plane within our approximations. Hence
 \begin{equation}
 -\frac{\partial}{\partial z}\left(D\frac{\partial \phi^S}{\partial z}\right)\simeq 2h\,\Gamma_{P\to S}\,\phi^P\,\delta(z)-2h\,\Gamma_{S}\,\phi^S\delta(z)\,,
\end{equation}
which is formally analogous to eq.~\eqref{slab1P}; the solution is thus
\begin{equation}
\frac{\phi_0^S(p)}{\phi_0^P(p)}= \Gamma_{P\to S}\tau_{\rm eff, S}(p)\,\label{secovprim},
\end{equation}
where 
 \begin{equation}
\tau_{\rm eff, S}^{-1}(p)=\tau_d^{-1}(p)+\Gamma_{S}\,.
\end{equation}
Note that the RHS of eq.~(\ref{secovprim}) depends on  the diffusion timescale $\tau_d$,  the target density in the ISM which can be inferred otherwise, and quantities that are in principle measurable in the lab, while the LHS comes from CR measurements at the Earth. This allows one to deduce $\tau_d$. This is the essential method through which one can break the degeneracy between source parameters and propagation ones, allowing one to infer source spectra via e.g. eq.~(\ref{slab3P}). The uncertainties with which hadronic cross-sections are known limits  however the precision with which propagation parameters can be inferred, even in this very simple setup, and do introduce some correlation between ``hadronic'' parameters and diffusion ones. These uncertainties are nowadays determining the ultimate capacity to test propagation frameworks, and must be reduced in the future if one is to fully exploit the high-precision data currently collected by experiments like AMS-02~\footnote{\texttt{https://ams02.space}}.

\newpage
\acknowledgments
These notes will appear in “Foundations of Cosmic Ray Astrophysics”, Proceedings of the International School of Physics "Enrico Fermi", Course 208, Varenna, 24 - 29 June 2022, edited by F. Aharonian, E. Amato, and P. Blasi. The author thanks the organisers for their kind invitation to lecturing in such a prestigious and fascinating location, where history and science meet, and natural and artistic beauty blend. 

\appendix
\section{Rutherford scattering}\label{Ruthscatt}
Rutherford scattering can be defined as the classic problem of elastic scattering of charged particles by the Coulomb interaction: The target (of charge $Z_t$) does not recoil and is the center of the potential. The problem obeys cylindrical symmetry around the incoming projectile direction. 
Given the Coulomb potential of a target and the initial kinetic energy $K$ (at infinity) of the incoming particles of charge $Z_p$, the scattering angle $\theta$  is uniquely determined by the impact parameter $b$.
The  cross section relates the number of particles scattered into $\d\Omega=2\pi \sin \theta \d \theta$  to the number of particles with associated impact parameters between $b$ and $b+\d b$,
\begin{equation}
2\pi b\d b= \frac{\d \sigma}{\d \Omega}\d \Omega \Longrightarrow \frac{\d \sigma}{\d \Omega}=\frac{b}{\sin \theta}\frac{\d b}{\d \theta}\,.
\end{equation}
Solving the EoM in polar coordinates with the target at the center ($r=0$) 
\begin{equation}
\frac{\d^2u }{\d \theta^2}+u=-\frac{Z_p Z_t e^2}{2\,K\,b^2}\,,\:\:\: ({\rm Binet}\:{\rm equation})
\end{equation}
with $u=1/r$ subject to boundary conditions $u\to 0$, $r\sin \theta\to b$  when $\theta\to \pi$, one derives
$b(\theta)=Z_p Z_t e^2\cot(\theta/2) /(2 K)$. Using that, one infers
\begin{equation}
\frac{\d \sigma}{\d \Omega}=\frac{Z_p^2 Z_t^2 \alpha^2}{16 K^2 \sin^4 \theta/2}\,.
\end{equation}
The change of momentum $\Delta p = 2 Z_p^2 e^2/(b\,v)$ is thus associated to the change of kinetic energy $W= (\Delta p)^2/(2m_e)= 2Z_p^2e^4/(b^2\,v^2\,m_e)$, which allows one  to compute the stopping power  via eq.~(\ref{stpower}). 
\section{Maximum energy transfer in 2-body scattering}\label{WmaxApp}
Here we derive the maximum energy transfer in a 2-body scattering of a particle of mass $M$ and momentum {\bf p} on a target particle (eg. an atomic electron) that is at rest (Lab frame), of mass $m_e$. For simplicity, below we will indicate moduli of 3-momenta in Italic, i.e. $|{\bf p}|=p$. 
Conservation of energy before and after the scattering requires (a prime denotes the target electron outgoing momentum, and double prime the projectile outgoing  momentum):
\begin{equation}
\sqrt{p^{2} +M^{2} }+m_{e} =\sqrt{p^{\prime \prime 2} +M^{2} }+\sqrt{p'^{2} +m_e^{2}} \Rightarrow \sqrt{p^{2} +M^{2} } =\sqrt{p^{\prime \prime 2} +M^{2} }+W\label{econs}\,,
\end{equation}
where the latter follows from the definition of kinetic energy for the electron, $W=E-m_e$, and the relativistic dispersion relation
\begin{equation}
p'^2=(W+m_{e} )^2-m_e^2\label{disp}\,.
\end{equation}
We can rewrite eq.~(\ref{econs})  isolating $p''^2$ as
\begin{equation}
p^{\prime \prime 2}  =p^{2} +W^2-2W\sqrt{p^{2} +M^{2} } \label{econs2}\,.
\end{equation}

The momentum conservation implies ($\theta$ being the angle between the scattered electron and the CR  incoming direction)
\begin{equation}
{\bf p}''={\bf p}-{\bf p}' \Rightarrow {p}''^2={ p}^2+{ p}'^2-2p\,p'\cos \theta\label{pcons}\,.
\end{equation}

Now, we can replace ${ p}'$ from eq.~(\ref{disp}) and ${p}''$ from eq.~(\ref{econs2}) in eq.~(\ref{pcons}) getting
\begin{eqnarray}
&&\nonumber-2W\sqrt{p^{2} +M^{2} }=2Wm_{e} -2p\sqrt{W^2+2m_{e}W}\cos \theta\\
&&\nonumber\Rightarrow\sqrt{1+2m_{e}/W}=\frac{m_{e} +\sqrt{p^{2} +M^{2} }}{p\cos\theta}\\
&&\Rightarrow W=\frac{2m_{e} p^2\cos^2\theta}{(m_{e} +\sqrt{p^{2} +M^{2} })^2-p^2\cos^2\theta}
\end{eqnarray}
The maximum energy transfer, i.e. the maximum of $W$, is obtained for $\cos \theta=1$ (head-on collision, the electron recoils along the incoming direction of the particle):
\begin{equation}
W_{\rm max}=\frac{2m_{e} p^2}{m_{e}^2 +M^2+2m_e\sqrt{p^{2} +M^{2} }}=\frac{2m_{e} \beta^2\gamma^2}{1+2\frac{m_e}{M}\gamma+\frac{m_{e}^2}{M^2} }\to 2m_e\beta^2\gamma^2
\label{Wmax}
\end{equation}
where the last limit holds when $M\gg m_e$. Eq.~(\ref{Wmax}) is the relation presented in the main text.

\section{From QED to Rutherford}\label{QEDtoRuth}
The notions recalled here are treated in an extensive way in any introductory  particle physics and quantum field theory textbook, such as~\cite{HM84,PS95,S13}.
The interaction probability in a two-body process ${1}+{2}\to  f_1+\ldots f_n$ is quantified via the cross-section $\sigma$. It can be written in a manifestly Lorentz-invariant form as
\begin{equation}
\d \sigma=\frac{|\mathcal{M}|^{2}}{F}\d Q\,, \label{genxsec}
\end{equation}
 where the phase-space factor $\d Q$ (imposing 4-momentum conservation) writes
\begin{equation}
\d Q=(2 \pi)^{4} \delta^{4}\left(p_{1}+p_{2}-\sum_{f}p_{f}\right) \prod_{f=f_1}^{f_n} \frac{\d^{3} p_{f}}{(2 \pi)^{3} 2 E_{f}}\,,
\end{equation}
while the flux factor $F$ writes
\begin{equation}
F=4\sqrt{(p_{1}\cdot p_{2})^2-m_{1}^2m_{2}^2}=2 E_{1} 2 E_{2} |v_{1}-v_{2}|\,.
\end{equation}
Finally, ${\cal M}$ is the initial to final transition matrix element (computed according to standard Feynman rules). If polarisation is not measured, as it is typically the case, $\left|{\cal M}\right|^2$ in eq.~(\ref{genxsec})  is intended as a sum over all final spin  states and an average over initial ones and denoted as $\overline{|\mathcal{M}|^{2}}$. By the way, the differential decay rate of a particle $1$ also writes in the form given by eq.~(\ref{genxsec}), with the flux factor however simply reducing to $2E_1$.

If the final state is made of two particles, 3 and 4, one has
\begin{equation}
\d \sigma=\frac{\overline{|\mathcal{M}|^{2}}}{8\pi^2 F} \beta_3p_3\d p_{3}\d \Omega_3\frac{\d^{3} p_4}{2 E_4}\delta^{4}\left(p_{1}+p_{2}-p_{3}-p_4\right)\,,
\end{equation}
an expression one can integrate over the ``uninteresting'' variables to get the differential (or total) cross section of interest. In doing that,  the following identity is often useful
\begin{equation}
\int\frac{\d^{3} p'}{2 E'}\delta^{4}\left(P-p'\right)=\int\d^{4} p'\delta^{4}\left(P-p'\right)\Theta(E')\delta(p'^2-m'^2)\,.\label{deltaid}
\end{equation}
Further manipulations require expressing $\overline{|\mathcal{M}|^{2}}$ and $F$ in a specific frame. In the CoM frame, for instance, it is a textbook exercise to prove that
\begin{equation}
\left.\frac{\d \sigma}{\d \Omega}\right|_{\rm CoM}=\frac{\overline{|{\cal M|}^2}}{64 \pi^2 E_{\rm CoM}^2}\frac{p'}{p}\,,\label{CoMxsecRut}
\end{equation}
where $p$, $p'$ are the initial and final momentum of either of the scattered particles.  
In the Lab frame, one has $F=4\beta\,E\,M$, with $E$ ($\beta$) the energy (respectively, speed) of the projectile and $M$ the mass of the target. 

Let us express $\overline{|\mathcal{M}|^{2}}$ for the case of the scattering $e+\mu\to e+\mu$, indicating with a prime (unprimed) symbol the final (initial) momenta, using $E, p, m$ for the electron energy, momentum, and mass and $k, M$ for the muon momentum and mass, respectively. This is clearly a proxy also for the electron-proton scattering, if we neglect the non-elementary nature of the proton and focus on it as an electromagnetically interacting spinor, much more massive than the electron. Also, here $_{\rm SI}$ denotes the SI charge, to comply with the standard QFT convention, so that $e_{\rm SI}^2=4\pi e^2=4\pi \alpha$.  One has
\begin{equation}
\overline{|\mathcal{M}|^{2}}=\frac{8e_{\rm SI}^4}{q^4}\left[(k'\cdot p')(k\cdot p)+(k\cdot p')(k'\cdot p)-m^2(k'\cdot k)-M^2(p'\cdot p)+2m^2M^2\right]\,.\label{M2emu}
\end{equation}
If we specify eq.~(\ref{M2emu}) to the frame where the heavy particle is initially at rest, in the relativistic limit for the electron, one has $q^2=-4E'E\sin^2\vartheta/2$, $\vartheta$ being the angle of the scattered electron with respect to its incoming momentum, 
\begin{equation}
\overline{|\mathcal{M}|^{2}}\to 
\frac{8e_{\rm SI}^4}{q^4}2M^2E'E\left\{\cos^2\vartheta/2-\frac{q^2}{2M^2}\sin^2\vartheta/2\right\}\,,
\end{equation}
and eq.~(\ref{deltaid}) applied to the muon momentum reduces to $(2M{\cal A})^{-1}\delta(E'-E/{\cal A})$
where ${\cal A}=1+(2E/M)\sin^2\theta/2$. This is equivalent to the kinematical relation following from conservation of energy and momentum, used to express the final electron energy in terms of the initial one. When combining these relations, after integration over $E'$ one obtains
\begin{equation}
\left.\frac{\d \sigma}{\d \Omega}\right|_{\rm Lab}=\frac{\alpha^2\cos^2\theta/2}{4E^2\sin^4\vartheta/2}\frac{1}{\left[1+\frac{2E}{M}\sin^2\vartheta/2\right]}\left\{1-\frac{q^2}{2M^2}\tan^2\vartheta/2\right\}\label{emuxsec}\,.
\end{equation}
Eq.~(\ref{emuxsec}) beautifully illustrates a number of notions: First of all, in the limit $M^2\to \infty$ only the first factor survives, i.e. one recovers Mott scattering, namely the Rutherford cross-section~\footnote{Note that Rutherford cross section can also be recovered from eq.~(\ref{CoMxsecRut}), if we use $E_{\rm CoM}\to M$, $p=p'$, and $\overline{|{\cal M}|^2}\to (2 m)^2(2M)^2/q^4$, with the $2m$ being the non-relativistic normalisations of the fermionic states.} corrected by the $\cos^2\vartheta/2$ factor which accounts for the spinorial nature of the electron.  The factor in brackets, going to $1$ in the limit  $M^2\to \infty$, accounts instead for the recoil of the target. Finally, the factor in curly brackets, also going to $1$ in the limit  $M^2\to \infty$, accounts for the spin of the target: The sub-leading term suppressed by $q^2/M^2$ accounts for the  scattering via the magnetic moment. This is analogous to the effect of the electron spin: The term $\cos^2\vartheta/2$, truly given by $1-\beta^2\sin^2 \vartheta$ without relativistic approximation, reflects the scattering on the electric charge $-e_{\rm SI}$ (the ``1'' term) and the magnetic moment $-e_{\rm SI}/2m$, respectively.  This can  be seen explicitly via the so-called Gordon decomposition, amounting to rewrite the spinor current $\bar{u}\gamma^\mu u$ as $(2m)^{-1}\bar{u}[(p'+p)^\mu+i\sigma^{\mu\nu}(p'-p)_\nu] u$. The first piece in this decomposition, if used to compute spinorial amplitudes, leads to the same results as a ``scalar'' electron with vertex $ie_{\rm SI}(p+p')^\mu$ instead of $ie_{\rm SI}\gamma^\mu$, and no spinor factors $u$ associated to the particle currents.

\section{More detailed derivation of the synchrotron spectrum.}\label{synchdet}
A full treatment of this topic can be found in~\cite{RLrpap91} and it is only sketched below. 
As discussed around eq.~(\ref{Poynting}), the power per unit solid angle carried by waves across the surface normal to their propagation direction is given by the Poynting vector. The emitted spectrum (energy per unit frequency and solid angle) of an accelerated charge can be thus expressed as Fourier transform of the modulus of eq.~(\ref{Poynting}), leading to
\begin{equation}
\frac{\d W}{\d \omega \d \Omega}=\frac{1}{4\pi^2}\left|\int \Re{\bf E}(t)\, e^{i\omega t}\d t\right|^{2}=\frac{e^{2} \omega^{2}}{4 \pi^{2} }\left|\int_{-\infty}^{\infty} \hat{n} \times(\hat{n} \times \vec{\beta}) e^{i \omega\left(t^{\prime}- \hat{n} \cdot \vec{r}\left(t^{\prime}\right)\right)} \d t^{\prime}\right|^{2}\label{LW1}\,,
\end{equation}
where the expression at the RHS comes from the differentiation of the  Li\'enard-Wechart  potentials
(whose details can be found in Sec. 14.1 of~\cite{Jackson}),  $\hat{n}$ is the  line of sight unit vector, $\vec\beta$ is the particle velocity and $t'$ is the retarded time accounting for the movement of the source, whose trajectory is described by $\vec{r}\left(t^{\prime}\right)$. This expression follows from changing the integration variable to $t'$ and assuming the observer far from the source. We remind the reader that we denote the pitch angle (angle between particle velocity and magnetic field) with $\zeta$ and that
the particle trajectory is characterised by the Larmor radius $r_L=\gamma r_g$.

Let us assume that the particle is at the origin of the reference system at $t'=0$,  $\vec{r}\left(t^{\prime}=0\right)=0$, with the magnetic field along $z$ and the velocity along $x$, denoting with $\theta$ the angle between the velocity and the line of sight.  As evident from eq.~(\ref{LW1}), the emission is perpendicular to the line of sight, and it is conveniently decomposed in one component parallel ($\hat{\epsilon}_{\|}$) and one perpendicular ($\hat{\epsilon}_{\perp} $) to the magnetic field direction. Since after a time $t'$ the particle has covered an angle $vt'/r_L$, we have
\begin{equation}
\hat{n} \times(\hat{n} \times \vec{\beta})=-\hat{\epsilon}_{\perp} \sin \left(\frac{v t^{\prime}}{r_L}\right)+\hat{\epsilon}_{\|} \cos \left(\frac{v t^{\prime}}{r_L}\right) \sin \theta\label{LW2}
\end{equation}
and
\begin{equation}
t'-\hat{n}\cdot \vec{r}(t')=t'-r_L\cos\theta\sin \left(\frac{v t^{\prime}}{r_L}\right)\,.
\end{equation}
Hence we can rewrite eq.~(\ref{LW1}) as
\begin{equation}
\frac{\d W}{\d \omega \d \Omega}=\frac{\alpha \omega^{2}}{4 \pi^{2} }
\left|
\int_{-\infty}^{\infty}\d t' 
\left[
\hat{\epsilon}_{\|}\sin\theta\cos\left(\frac{vt'}{r_L}\right)-\hat\epsilon_{\perp}\sin\left(\frac{vt'}{r_L}\right)
\right]
e^{i\omega \left(t'-r_L\cos\theta\sin(\frac{vt'}{r_L})\right)}
\right|^2\,.
\end{equation}
It makes sense to split this expression as follows
\begin{equation}
\frac{\d W}{\d \omega \d \Omega}=\left(\frac{\d W}{\d \omega \d \Omega}\right)_{\perp}+\left(\frac{\d W}{\d \omega \d \Omega}\right)_{\|}\,.
\end{equation}

Let us introduce $\theta^2_\gamma\equiv 1+\theta^2\gamma^2$, which turns out to control the angular dependence, the new variable of integration $y=\gamma t'/(a\theta_\gamma)$, and the new variable $\eta\equiv \omega r_L \theta_\gamma^3/ (3\gamma^3)$  which retains the complete frequency-dependence. 
For relativistic particles $\beta\approx 1 $ ($\gamma \gg1$) the observer receives the signal only when $vt'/r_L\approx0$
and $\theta\approx 0$ (remember Fig.s~\ref{fig:beaming} and \ref{fig:beaming2}), so that we can expand the trigonometric functions accordingly, and also consider $\eta\approx \eta(\theta=0)=\omega/(2\omega_c)$, where $\omega_c\equiv 3\gamma^2 \omega_g\sin\zeta/2$.  Then eq.~(\ref{LW2}) yields

\begin{equation}
\left(\frac{\d W}{\d \omega \d \Omega}\right)_{\perp}=\frac{\alpha \omega^{2}}{4 \pi^{2} }\left(\frac{r_L\theta_\gamma^2}{\gamma^2}\right)^2\left| \int_{-\infty}^{\infty}\d y\,y\,e^{\frac{3}{2} i \eta \left(y+\frac{y^3}{3}\right)} \right|^2
=\frac{\alpha \omega^{2}}{3 \pi^{2} }\left(\frac{r_L\theta_\gamma^2}{\gamma^2}\right)^2K_{2/3}^2(\eta)\,,
\end{equation}

\begin{equation}
\left(\frac{\d W}{\d \omega \d \Omega}\right)_{\|}=\frac{\alpha \omega^{2}\theta^2}{4 \pi^{2} }\left(\frac{r_L\theta_\gamma}{\gamma}\right)^2\left| \int_{-\infty}^{\infty}\d y\,e^{\frac{3}{2} i \eta \left(y+\frac{y^3}{3}\right)} \right|^2=\frac{\alpha \omega^{2}}{3 \pi^{2} }\left(\frac{r_L\theta_\gamma}{\gamma}\right)^2K_{1/3}^2(\eta)\,,
\end{equation}
where $K_i(x)$ are  the modified Bessel functions of order $i$.

We can now integrate these relations over the solid angle, obtaining in this way the
  energy emitted per unit frequency for a complete orbit on the plane normal to the magnetic field.
During one such orbit the emitted radiation is almost completely confined to the solid angle lying within an angle $1/\gamma$ of a cone of aperture twice the pitch angle $\zeta$. Thus $\d\Omega=2\pi\sin\zeta \d\theta$ and 
\begin{equation}
\left(\frac{\d W}{\d \omega \d \Omega}\right)_{\perp}\simeq\frac{2 \alpha \omega^{2}r_L^2\sin\zeta}{3 \pi \gamma^{4} } 
 \int_{-\infty}^{\infty}\d\theta\, \theta_\gamma^4\,K^2_{2/3}(\eta)= \frac{\sqrt{3}\alpha\gamma \sin\zeta}{2}[F(x)+G(x)]\,,
\end{equation}
\begin{equation}
\left(\frac{\d W}{\d \omega \d \Omega}\right)_{\|}\simeq\frac{2 \alpha\omega^{2}r_L^2\sin\zeta}{3 \pi \gamma^{2} } 
 \int_{-\infty}^{\infty}\d\theta\, \theta_\gamma^2\,K^2_{1/3}(\eta)= \frac{\sqrt{3}\alpha\gamma \sin\zeta}{2}[F(x)-G(x)]\,,
\end{equation}
where $x\equiv \omega/\omega_c$,  we have extended the integrals to infinity for analytical ease, since anyway the integrands are non-vanishing only in a small angular interval, and defined
\begin{equation}
F(x)\equiv x\int_x^\infty K_{5/3}(\xi)\d \xi\,,\:\:\:G(x)\equiv xK_{2/3}(x)\,.
\end{equation}

The total emitted energy per unit frequency is the sum of the two contributions above,
hence given by
\begin{equation}
\left(\frac{\d W}{\d \omega \d \Omega}\right)=\frac{\sqrt{3}\alpha\gamma \sin\zeta}{2}F(x)\,.
\end{equation}
The function $F(x)$ scales as $x^{1/3}$ at $x\ll 1$, and as $x^{1/2}e^{-x}$ at $x\gg 1$. The total power per unit frequency can be obtained by dividing the energy by the giration period of the particle, $2\pi \gamma/\omega_g=\gamma/\nu_g$, so that:
\begin{equation}
P(\omega)=\sqrt{3}\alpha \nu_g \sin\zeta F(x)\,.
\end{equation}
Note that this spectrum peaks around $\omega_c\sim \gamma^2 \omega_g$, as argued with an heuristic argument in the main text.


\begin{thebibliography}{0}

\bibitem{Rossi} B. Rossi, ``High Energy Particles'', Prentice-Hall, Inc., Englewood Cliffs, NJ, (1952).

\bibitem{Jackson}
J. D. Jackson, ``Classical Electrodynamics'', 3$^{\rm rd}$ edition, John Wiley \& Sons (1999).


\bibitem{John:2022asa}
I.~John and T.~Linden,
``Pulsars Do Not Produce Sharp Features in the Cosmic-Ray Electron and Positron Spectra,''
Phys. Rev. D \textbf{107} (2023) no.10, 103021
[arXiv:2206.04699 [astro-ph.HE]].

\bibitem{Esmaeili:2022cpz}
A.~Esmaeili, A.~Capanema, A.~Esmaili and P.~D.~Serpico,
``Ultrahigh energy neutrinos from high-redshift electromagnetic cascades,''
Phys. Rev. D \textbf{106} (2022) no.12, 123016
[arXiv:2208.06440 [hep-ph]].



\bibitem{LHCForwardPhysicsWorkingGroup:2016ote}
K.~Akiba \textit{et al.} [LHC Forward Physics Working Group],
``LHC Forward Physics,''
J. Phys. G \textbf{43}, 110201 (2016)
[arXiv:1611.05079 [hep-ph]].

\bibitem{1966PhRv..150.1088A} P.~B. Abraham, K.~A. Brunstein, T.~L. Cline,\, ``Production of Low-Energy Cosmic-Ray Electrons'', Physical Review, 150, 1088 (1966). 


\bibitem{1976ApJ...206..312O} C.~D.~Orth, A. Buffington, ``Secondary cosmic-ray $e^\pm$ from 1 to 100 GeV in the upper atmosphere and interstellar space, and interpretation of a recent e$^{+}$ flux measurement,''
Astrophys. J.  \textbf{206}, 312 (1976). 

\bibitem{Maurin:2001yxa}
D.~Maurin,
``Propagation des rayons cosmiques dans un mod\`ele de diffusion: une nouvelle estimation des param\`etres de diffusion et du flux d'antiprotons secondaires,''
PhD thesis Savoie U (2001).


\bibitem{1947ZNatA...2..133M}G. Moli{\`e}re, ``Theorie der Streuung schneller geladener Teilchen I. Einzelstreuung am abgeschirmten Coulomb-Feld'', Zeitschrift Naturforschung Teil A, 2, 133 (1947) 
\bibitem{1948ZNatA...3...78M} G. Moli{\`e}re, ``Theorie der Streuung schneller geladener Teilchen II. Mehrfach- und Vielfachstreuung'', Zeitschrift Naturforschung Teil A, 3, 78 (1948). 


\bibitem{Bethe:1953va}
H.~A.~Bethe,
``Moli\`ere's theory of multiple scattering,''
Phys. Rev. \textbf{89}, 1256-1266 (1953)

\bibitem{Mannheim:1994sv}
K.~Mannheim and R.~Schlickeiser,
``Interactions of Cosmic Ray Nuclei,''
Astron. Astrophys. \textbf{286}, 983-996 (1994)
[astro-ph/9402042].


\bibitem{Strong:1998pw}
A.~W.~Strong and I.~V.~Moskalenko,
``Propagation of cosmic-ray nucleons in the galaxy,''
Astrophys. J. \textbf{509} (1998), 212-228.

\bibitem{Evoli:2016xgn}
C.~Evoli {\it et al.}, 
``Cosmic-ray propagation with $\small{DRAGON2}$: I. numerical solver and astrophysical ingredients,''
JCAP \textbf{02} (2017), 015
[arXiv:1607.07886].



\bibitem{Uehling:1954wp}
E.~A.~Uehling,
``Penetration of heavy charged particles in matter,''
Ann. Rev. Nucl. Part. Sci. \textbf{4} (1954), 315-350.


\bibitem{Ahlen:1980xr}
S.~P.~Ahlen,
``Theoretical and experimental aspects of the energy loss of relativistic heavily ionizing particles,''
Rev. Mod. Phys. \textbf{52} (1980), 121-173
[erratum: ibidem, pages 653-653].


\bibitem{ParticleDataGroup:2020ssz}
P.~A.~Zyla \textit{et al.} [Particle Data Group],
``Review of Particle Physics,''
PTEP \textbf{2020}, no.8, 083C01 (2020).

\bibitem{Moeller31} C. M{\o}ller, ``\"Uber die Wechselwirkung von Zwei Elektronen'', Zeitschrift fur Physik, 70, 786 (1931).
 
\bibitem{1932ZPhy...77..296B}H. Bethe, E. Fermi, ``\"Uber die Wechselwirkung von Zwei Elektronen'', Zeitschrift fur Physik, 77, 296 (1932).

\bibitem{1932RvMP....4...87F} E. Fermi, ``Quantum Theory of Radiation'', Reviews of Modern Physics, 4, 87 (1932).

\bibitem{Heitler:1936jqw}
W.~Heitler,
``The quantum theory of radiation,''  International Series of Monographs on Physics, Oxford (1936).


\bibitem{Schweber02} S. Schweber, ``Enrico Fermi and Quantum Electrodynamics, 1929–32'' Physics Today 55, 6, 31 (2002). 


\bibitem{Bhabha:1936zz}
H.~J.~Bhabha,
``The scattering of positrons by electrons with exchange on Dirac's theory of the positron,''
Proc. Roy. Soc. Lond. A \textbf{154}, 195-206 (1936)

\bibitem{Padovani:2009bj}
M.~Padovani, D.~Galli and A.~E.~Glassgold,
``Cosmic-ray ionization of molecular clouds,''
Astron. Astrophys. \textbf{501}, 619 (2009)
[arXiv:0904.4149 [astro-ph.SR]].

\bibitem{Gabici:2022nac}
S.~Gabici,
``Low energy cosmic rays: Regulators of the dense interstellar medium,''
Astron. Astrophys. Rev. \textbf{30} (2022) no.1, 4.
[arXiv:2203.14620 [astro-ph.HE]]

\bibitem{BH34} H. Bethe and W. Heitler, ``On the Stopping of Fast Particles and on the Creation of Positive Electrons,'' Proc.Roy. Soc., {\bf 146}, 83 (1934).


\bibitem{RLrpap91} G. B. Rybicki and A. P. Lightman, ``Radiative Processes in Astrophysics'', Wiley-VCH (1991).


\bibitem{Ghisellini:2012hs}
G.~Ghisellini,
``Radiative Processes in High Energy Astrophysics,''
Lect. Notes Phys. \textbf{873}, 1-147  (2013)
[arXiv:1202.5949].


  \bibitem{Blumenthal:1970gc}
G.~R.~Blumenthal and R.~J.~Gould,
``Bremsstrahlung, synchrotron radiation, and compton scattering of high-energy electrons traversing dilute gases,''
Rev. Mod. Phys. \textbf{42}, 237-270 (1970).

\bibitem{Gouldbook}
R. J. Gould, ``Electromagnetic Processes (Princeton Series in Astrophysics)'', Princeton University Press (2020).



\bibitem{Akahori:2017lhe} 
  T.~Akahori {\it et al.},
  ``Cosmic Magnetism in Centimeter and Meter Wavelength Radio Astronomy,''
  Publ.\ Astron.\ Soc.\ Jap.\  {\bf 70}, no. 1, R2 (2018)
  [arXiv:1709.02072].
  
  
\bibitem{Ajello:2018sxm} 
  S.~Abdollahi {\it et al.} [Fermi-LAT Collaboration],
``A gamma-ray determination of the Universe's star formation history,''
  Science {\bf 362}, no. 6418, 1031 (2018)
  [arXiv:1812.01031].
  
\bibitem{KN29}
O. Klein, T. Nishina, ``\"Uber die Streuung von Strahlung durch freie Elektronen nach der neuen relativistischen Quantendynamik von Dirac,'' Zeitschrift fur Physik {\bf 52}, 853-868(1929).

 \bibitem{WW00}
M. S. Zolotorev, K. T. McDonald, ``Classical Radiation Processes in the Weizsacker-Williams Approximation,'' arXiv:physics/0003096.


\bibitem{Ressler:2014gza}
S.~M.~Ressler {\it et al.}, 
``Magnetic-Field Amplification in the Thin X-ray Rims of SN 1006,''
Astrophys. J. \textbf{790}, no.2, 85 (2014)
[arXiv:1406.3630 [astro-ph.HE]].

\bibitem{Bergstrom:1999jc}
L.~Bergstrom, J.~Edsjo and P.~Ullio,
``Cosmic anti-protons as a probe for supersymmetric dark matter?,''
Astrophys. J. \textbf{526} (1999), 215-235
[arXiv:astro-ph/9902012 [astro-ph]].


\bibitem{Varenna53} G. Puppi (editor), ``Questioni relative alla rivelazione delle particelle elementari, con particolare riguardo alla radiazione cosmica'', Proceedings of the International School of Physics ``Enrico Fermi'', Vol. I (1953).

\bibitem{Cronin:2011zz}
J.~W.~Cronin, ``The 1953 cosmic ray conference at Bagneres de Bigorre: The Birth of sub atomic physics,''
Eur. Phys. J. H \textbf{36}, 183-201 (2011)
[arXiv:1111.5338 [physics.hist-ph]].



\bibitem{Varenna54} G. Puppi (editor), ``Questioni relative alla rivelazione delle particelle elementari, e alle loro interazioni con particolare riguardo alle particelle artificialmente prodotte ed accelerate'', Proceedings of the International School of Physics ``Enrico Fermi'', Vol. II (1954).


\bibitem{Letaw:1983}
J.~R.~Letaw, R.~Silberberg, C.~H.~Tsao, 
``Proton-nucleus total inelastic cross sections - an empirical formula for E greater than 10 MeV,''
 The Astrophys. J. Suppl. Series {\bf 51}, 271-276, 1983.
 
 
\bibitem{Mazziotta:2015uba}
M.~N.~Mazziotta, F.~Cerutti, A.~Ferrari, D.~Gaggero, F.~Loparco and P.~R.~Sala,
``Production of secondary particles and nuclei in cosmic rays collisions with the interstellar gas using the FLUKA code,''
Astropart. Phys. \textbf{81}, 21-38 (2016)
[arXiv:1510.04623 [astro-ph.HE]].

\bibitem{Krakau:2015bea}
S.~Krakau and R.~Schlickeiser,
``Pion Production Momentum Loss of Cosmic ray Hadrons,''
Astrophys. J. \textbf{802} (2015) no.2, 114.



\bibitem{Kelner:2006tc}
S.~R.~Kelner, F.~A.~Aharonian and V.~V.~Bugayov,
``Energy spectra of gamma-rays, electrons and neutrinos produced at proton-proton interactions in the very high energy regime,''
Phys. Rev. D \textbf{74} (2006), 034018
[erratum: Phys. Rev. D \textbf{79} (2009), 039901]
[astro-ph/0606058].


\bibitem{Evoli:2017vim}
C.~Evoli {\it et al.}, 
``Cosmic-ray propagation with DRAGON2: II. Nuclear interactions with the interstellar gas,''
JCAP \textbf{07} (2018), 006
[arXiv:1711.09616].	

\bibitem{Genolini:2018ekk}
Y.~Genolini, D.~Maurin, I.~V.~Moskalenko and M.~Unger,
``Current status and desired precision of the isotopic production cross sections relevant to astrophysics of cosmic rays: Li, Be, B, C, and N,''
Phys. Rev. C \textbf{98} (2018) no.3, 034611
[arXiv:1803.04686].	



\bibitem{Longair}
M.~S.~Longair,
``High-energy astrophysics'', 3$^{\rm rd}$ Edition, Cambridge University Press (2011).




\bibitem{Ramaty:1979}
R.~Ramaty, B.~Kozlovsky and R.~E.~Lingenfelter,
``Nuclear gamma-rays from energetic particle interactions.''
Astrophys. J. Supp. \textbf{40}, 487-526 (1979)


\bibitem{AGILE:2011tzq}
A.~Giuliani \textit{et al.} [AGILE],
``Neutral pion emission from accelerated protons in the supernova remnant W44,''
Astrophys. J. Lett. \textbf{742} (2011), L30
[arXiv:1111.4868].

\bibitem{Fermi-LAT:2013iui}
M.~Ackermann \textit{et al.} [Fermi-LAT],
``Detection of the Characteristic Pion-Decay Signature in Supernova Remnants,''
Science \textbf{339} (2013), 807
[arXiv:1302.3307].


\bibitem{Kafexhiu:2014cua} 
  E.~Kafexhiu, F.~Aharonian, A.~M.~Taylor and G.~S.~Vila,
  ``Parametrization of gamma-ray production cross-sections for pp interactions in a broad proton energy range from the kinematic threshold to PeV energies,''
  Phys.\ Rev.\ D {\bf 90}, no. 12, 123014 (2014)
  [arXiv:1406.7369].     
  
  
  
\bibitem{dermer1986} C.D.~Dermer, ``Binary Collision Rates of Relativistic Thermal Plasmas. II. Spectra,'' Astrophysical Journal 307  47 (1986).


\bibitem{Halzen:2021ynx}
F.~Halzen,
``The observation of high-energy neutrinos from the cosmos: Lessons learned for multimessenger astronomy,''
Int. J. Mod. Phys. D \textbf{31} (2022) no.03, 2230003
[arXiv:2110.01687].

\bibitem{Boudaud:2019efq}
M.~Boudaud {\it et al.}, 
``AMS-02 antiprotons' consistency with a secondary astrophysical origin,''
Phys. Rev. Res. \textbf{2}, no.2, 023022 (2020)
[arXiv:1906.07119].

\bibitem{diMauro:2014zea}
M.~di Mauro, F.~Donato, A.~Goudelis and P.~D.~Serpico,
``New evaluation of the antiproton production cross section for cosmic ray studies,''
Phys. Rev. D \textbf{90}, no.8, 085017 (2014)
[erratum: Phys. Rev. D \textbf{98}, no.4, 049901 (2018)]
[arXiv:1408.0288 [hep-ph]].


\bibitem{Lin:2016ezz}
S.~J.~Lin, X.~J.~Bi, J.~Feng, P.~F.~Yin and Z.~H.~Yu,
``Systematic study on the cosmic ray antiproton flux,''
Phys. Rev. D \textbf{96}, no.12, 123010 (2017)
[arXiv:1612.04001 [astro-ph.HE]].

\bibitem{Winkler:2017xor}
M.~W.~Winkler,
``Cosmic Ray Antiprotons at High Energies,''
JCAP \textbf{02}, 048 (2017)
[arXiv:1701.04866 [hep-ph]].


\bibitem{Donato:2017ywo}
F.~Donato, M.~Korsmeier and M.~Di Mauro,
``Prescriptions on antiproton cross section data for precise theoretical antiproton flux predictions,''
Phys. Rev. D \textbf{96}, no.4, 043007 (2017)
[arXiv:1704.03663 [astro-ph.HE]].


\bibitem{Korsmeier:2018gcy}
M.~Korsmeier, F.~Donato and M.~Di Mauro,
``Production cross sections of cosmic antiprotons in the light of new data from the NA61 and LHCb experiments,''
Phys. Rev. D \textbf{97}, no.10, 103019 (2018)
[arXiv:1802.03030 [astro-ph.HE]].






\bibitem{Calore:2022stf}
F.~Calore {\it et al.}, 
``AMS-02 antiprotons and dark matter: Trimmed hints and robust bounds,''
SciPost Phys. \textbf{12}, no.5, 163 (2022)
[arXiv:2202.03076 [hep-ph]].



\bibitem{Donato:1999gy}
F.~Donato, N.~Fornengo and P.~Salati,
``Anti-deuterons as a signature of supersymmetric dark matter,''
Phys. Rev. D \textbf{62}, 043003 (2000)
[hep-ph/9904481]


\bibitem{Cirelli:2014qia}
M.~Cirelli, N.~Fornengo, M.~Taoso and A.~Vittino,
``Anti-helium from Dark Matter annihilations,''
JHEP \textbf{08}, 009 (2014)
[arXiv:1401.4017].


\bibitem{Carlson:2014ssa}
E.~Carlson {\it et al.}, 
``Antihelium from Dark Matter,''
Phys. Rev. D \textbf{89}, no.7, 076005 (2014)
[arXiv:1401.2461].

\bibitem{Mori:2001dv}
K.~Mori, C.~J.~Hailey, E.~A.~Baltz, W.~W.~Craig, M.~Kamionkowski, W.~T.~Serber and P.~Ullio,
``A Novel antimatter detector based on x-ray deexcitation of exotic atoms,''
Astrophys. J. \textbf{566}, 604-616 (2002)
[arXiv:astro-ph/0109463 [astro-ph]].

\bibitem{Serksnyte:2022onw}
L.~\v{S}erk\v{s}nyt\.{e} {\it et al,}, 
``Reevaluation of the cosmic antideuteron flux from cosmic-ray interactions and from exotic sources,''
Phys. Rev. D \textbf{105} (2022) no.8, 083021
[arXiv:2201.00925 [astro-ph.HE]].


\bibitem{Shukla:2020bql}
A.~Shukla, A.~Datta, P.~von Doetinchem, D.~M.~Gomez-Coral and C.~Kanitz,
``Large-scale Simulations of Antihelium Production in Cosmic-ray Interactions,''
Phys. Rev. D \textbf{102}, no.6, 063004 (2020)
[arXiv:2006.12707].




\bibitem{AstroCR}
V. S Berezinskii {et al.} ``Astrophysics of cosmic rays'' (edited by V.L Ginzburg)	Amsterdam: North-Holland, 1990.

\bibitem{Berezinsky:2016feh}
V.~Berezinsky and O.~Kalashev,
``High energy electromagnetic cascades in extragalactic space: physics and features,''
Phys. Rev. D \textbf{94}, no.2, 023007 (2016)
[arXiv:1603.03989].



\bibitem{Capanema:2020oet}
A.~Capanema, A.~Esmaili and P.~D.~Serpico,
``Where do IceCube neutrinos come from? Hints from the diffuse gamma-ray flux,''
JCAP \textbf{02}, 037 (2021)
[arXiv:2007.07911].






\bibitem{HM84}
F. Halzen, A. D. Martin, ``Quarks and Leptons: An Introductory Course in Modern Particle Physics'', Wiley (1984).


\bibitem{PS95} 
M. E. Peskin, D. V. Schroeder, ``An Introduction To Quantum Field Theory'',  CRC Press (1995).


\bibitem{S13}
M. D. Schwartz, ``Quantum Field Theory and the Standard Model'', Cambridge Unv. Press (2013). 


\end{thebibliography}
\end{document}